\documentclass[11pt]{article}
\pdfoutput=1
\usepackage[utf8]{inputenc}
\usepackage[left=20mm,right=20mm,top=25mm,bottom=25mm]{geometry}
\usepackage[T1]{fontenc}

\usepackage[ngerman, main=english]{babel}

\usepackage[maxbibnames=10, maxcitenames=2, giveninits=true,
    natbib=true, bibencoding=utf8, isbn=false, url=false, backend=bibtex, style=authoryear-comp]{biblatex} 
\AtEveryBibitem{\clearfield{month}}
\AtEveryBibitem{\clearfield{day}}
\AtEveryBibitem{\clearlist{language}}
 
\usepackage[affil-it]{authblk}
\usepackage{pgfplots}
\pgfplotsset{compat=newest} 
\pgfplotsset{plot coordinates/math parser=false} 

\usepgfplotslibrary{external} 
\tikzsetexternalprefix{TikzFig/}
\pgfkeys{/pgf/number format/.cd,1000 sep={}} 

\pgfplotsset{every axis/.append style={
     legend style={nodes={font=\tiny,scale=1.2}} 
  		},
		every tick label/.append style={scale=0.8}, 
   }
\pgfplotsset{
  every axis plot/.append style={line width=0.75pt},
  every axis plot post/.append style={
    every mark/.append style={line width=0.4pt,mark size=1.3pt}
  }
}
\pgfplotsset{label style={font=\footnotesize}} 

\pgfplotsset{
        table/search path={tiks,tiks/ROM-signal-evals,tiks/ROM-signal,tiks/datasets},
    }

\usetikzlibrary{backgrounds} 
\pgfdeclarelayer{background}
\pgfdeclarelayer{foreground}
\pgfsetlayers{background,main,foreground} 
\usetikzlibrary{calc, patterns}
\usepgfplotslibrary{fillbetween}
\usetikzlibrary{arrows.meta,arrows}

\usepackage[autostyle]{csquotes}
\usepackage{microtype}

\usepackage{tabularx}
\usepackage{tabu}

\usepackage{booktabs}
\usepackage{varwidth} 
\usepackage{multirow}

\usepackage{adjustbox}

\usepackage{mathtools}
\usepackage{amssymb}
\usepackage{mathrsfs}
\usepackage{bm}
\usepackage{setspace}
\usepackage{xfp} 

\usepackage{hyperref}

\hypersetup{
    colorlinks=true,
    linkcolor=tud2a,
    filecolor=magenta,      
    urlcolor=tud2a,
    citecolor=tud2a,
    pdfpagemode=FullScreen,
    }

\usepackage{siunitx}
\DeclareSIUnit[quantity-product = ]\percent{\char`\%}

\usepackage{outlines} 

\usepackage{setspace}

\newcommand{\ItemSpacing}{2pt}%
\newcommand{\ParSpacing}{2pt}%

\usepackage{outlines}
\usepackage{enumitem}
\setenumerate[1]{itemsep={\ItemSpacing},parsep={\ParSpacing},label=\Roman*.}
\setenumerate[2]{itemsep={\ItemSpacing},parsep={\ParSpacing},label=\Alph*.}
\setenumerate[3]{itemsep={\ItemSpacing},parsep={\ParSpacing},label=\roman*.}
\setenumerate[4]{itemsep={\ItemSpacing},parsep={\ParSpacing},label=\alph*.}



\usepackage{lineno}

\usepackage{caption}
\usepackage{subcaption} 

\usepackage{algorithm}
\usepackage{algpseudocode}

\makeatletter
\newcommand\fs@nobottomruled{\def\@fs@cfont{\bfseries}\let\@fs@capt\floatc@ruled
  \def\@fs@pre{}
  \def\@fs@post{\kern2pt\hrule\relax}
  \def\@fs@mid{\kern2pt\hrule\kern2pt}%
  \let\@fs@iftopcapt\iftrue}
\makeatother

\floatstyle{nobottomruled}
\restylefloat{algorithm}

\newbibmacro{string+doiurlisbn}[1]{%
  \iffieldundef{doi}{%
    \iffieldundef{url}{#1}{%
      \href{\thefield{url}}{#1}%
    }
  }{%
    \href{http://doi.org/\thefield{doi}}{#1}%
  }%
}

\DeclareFieldFormat{title}{\usebibmacro{string+doiurlisbn}{\mkbibemph{#1}}}
\DeclareFieldFormat[article,incollection]{title}%
    {\usebibmacro{string+doiurlisbn}{\mkbibquote{#1}}}

\usepackage{acronym}

\usepackage{url}
\newcommand{\pages}[1]{\textbf{[#1]}}
\renewcommand{\pages}[1]{\textbf{}}
\newcommand{\side}[1]{\textit{(#1)}}
\renewcommand{\side}[1]{\textit{}}

\addto\extrasenglish{%
    \def\equationautorefname~#1\null{Eq. (#1)\null}
}

\definecolor{tud1a}{HTML}{5D85C3}
\definecolor{tud2atrue}{HTML}{009CDA}
\definecolor{tud3a}{HTML}{50B695}
\definecolor{tud3ai}{HTML}{009a67} 
\definecolor{tud4a}{HTML}{AFCC50}
\definecolor{tud5a}{HTML}{DDDF48}
\definecolor{tud6a}{HTML}{FFE05C}
\definecolor{tud6b}{HTML}{FDCA00}
\definecolor{tud8a}{HTML}{EE7A34}
\definecolor{tud8b}{HTML}{EC6500}
\definecolor{tud9a}{HTML}{E9503E}
\definecolor{tud11b}{HTML}{721085}
\definecolor{tud1b}{HTML}{005AA9} 
\definecolor{tud3b}{HTML}{009D81}
\definecolor{tud4b}{HTML}{99C000}
\definecolor{tud4c}{HTML}{7FAB16}
\definecolor{tud4d}{HTML}{6A8B22}
\definecolor{tud9b}{HTML}{E6001A}
\definecolor{tud9c}{HTML}{B90F22} 
\definecolor{tud10atab}{HTML}{E98174} 
\definecolor{sourcecode}{HTML}{E9E9E9}
\definecolor{mygreyold}{HTML}{C0C0C0}
\definecolor{mygrey}{RGB}{170, 170, 170}

\definecolor{tud2a}{HTML}{0072BD}%
\definecolor{tud10a}{HTML}{BE1622}
\definecolor{tud7b}{HTML}{F5A300}%

\definecolor{color11a}{RGB}{236, 188, 0}
\definecolor{color12a}{RGB}{50, 67, 121}
\definecolor{color13}{RGB}{222, 222, 222}
\definecolor{color21a}{RGB}{92, 134, 196}
\definecolor{color22a}{RGB}{249, 156, 0}
\definecolor{color23}{RGB}{222, 222, 222}
\definecolor{color31}{RGB}{222, 222, 222}
\definecolor{color32}{RGB}{222, 222, 222}
\definecolor{color33a}{RGB}{191, 60, 60}

\definecolor{tudtoven}{HTML}{B90F22} 
\definecolor{tudha}{HTML}{951169} 
\definecolor{tudgu}{HTML}{A60084} 
\definecolor{tudcw}{HTML}{7FAB16} 
\definecolor{tudli}{HTML}{CC4C03} 
\definecolor{tudrabeler}{HTML}{009CDA} 
\definecolor{tudmmodel}{HTML}{009D81} 

\definecolor{tudsolvedcases}{HTML}{00689D}
\definecolor{tudsolvedrate}{HTML}{951169} 

\definecolor{tudcpu}{HTML}{008877} 
\definecolor{tudspeedup}{HTML}{B90F22}

\newcommand{\overbar}[1]{\mkern 1.5mu\overline{\mkern-1.9mu#1\mkern-0.5mu}\mkern 0.5mu}

\newcommand{\disphat}[2][3mu]{\hat{#2\mkern#1}\mkern-#1}

\newcommand{\defeq}{\vcentcolon=}

\newcommand*\mapeq{\mathop{}\!E_\text{map}}
\newcommand*\rmseq{\mathop{}\!E_\text{rms}}
\newcommand*\maxeq{\mathop{}\!E_\text{max}}
\newcommand*\medeq{\mathop{}\!E_\text{med}}
\newcommand*\iqreq{\mathop{}\!E_\text{iqr}}

\newcommand*\mapeg{\overbar{E}_\text{map}}
\newcommand*\rmseg{\overbar{E}_\text{rms}}
\newcommand*\maxeg{\overbar{E}_\text{max}}
\newcommand*\medeg{\overbar{E}_\text{med}}
\newcommand*\iqreg{\overbar{E}_\text{iqr}}

\newcommand*\diff{\mathop{}\!\mathrm{d}}

\newcommand{\vect}[1]{\boldsymbol{#1}}

\newcommand{\Arch}{\operatorname{\mathit{A\kern-.06em r}}} 
\newcommand{\Biot}{\operatorname{\mathit{B\kern-.06em i}}} 
\newcommand{\Cauc}{\operatorname{\mathit{C\kern-.07em a}}} 
\newcommand{\Damk}{\operatorname{\mathit{D\kern-.06em a}}} 
\newcommand{\Eule}{\operatorname{\mathit{E\kern-.03em u}}} 
\newcommand{\Four}{\operatorname{\mathit{F\kern-.10em o}}} 
\newcommand{\Frou}{\operatorname{\mathit{F\kern-.07em r}}} 
\newcommand{\Gras}{\operatorname{\mathit{G\kern-.05em r}}} 
\newcommand{\Karl}{\operatorname{\mathit{K\kern-.11em a}}} 
\newcommand{\Knud}{\operatorname{\mathit{K\kern-.11em n}}} 
\newcommand{\Lewi}{\operatorname{\mathit{L\kern-.05em e}}} 
\newcommand{\Mach}{\operatorname{\mathit{M\kern-.10em a}}} 
\newcommand{\Nuss}{\operatorname{\mathit{N\kern-.20em u}}} 
\newcommand{\Pecl}{\operatorname{\mathit{P\kern-.14em e}}} 
\newcommand{\Pran}{\operatorname{\mathit{P\kern-.09em r}}} 
\newcommand{\Rayl}{\operatorname{\mathit{R\kern-.04em a}}} 
\newcommand{\Reyn}{\operatorname{\mathit{R\kern-.09em e}}} 
\newcommand{\Rich}{\operatorname{\mathit{R\kern-.06em i}}} 
\newcommand{\Schm}{\operatorname{\mathit{S\kern-.14em c}}} 
\newcommand{\Sher}{\operatorname{\mathit{S\kern-.07em h}}} 
\newcommand{\Stro}{\operatorname{\mathit{S\kern-.07em r}}} 
\newcommand{\Webe}{\operatorname{\mathit{W\kern-.14em e}}} 

\newcommand{\kelvin}{\si{\kelvin}}
\newcommand{\newton}{\si{\newton}}

\newcommand{\ha}{\operatorname{\mathit{H\kern-.25em a}}}
\newcommand{\gu}{\operatorname{\mathit{G\kern-.17em u}}}
\newcommand{\cw}{\operatorname{\mathit{C\kern-.2em w}}}

\graphicspath{{./fig/marker/}}

\title{
\textcolor{black}{TwinLab: a framework for data-efficient training of non-intrusive reduced-order models for digital twins}
}

\author[1,*]{Maximilian~Kannapinn}
\author[2]{Michael Schäfer}
\author[1]{Oliver~Weeger}
\affil[1]{\footnotesize Technical University of Darmstadt, Department of Mechanical Engineering,  Cyber-Physical Simulation}
\affil[2]{\footnotesize Technical University of Darmstadt, Department of Mechanical Engineering, Numerical Methods in Mechanical Engineering \protect \\
Dolivostr.~15, 64293 Darmstadt, Germany}
\affil[*]{\footnotesize Corresponding author, Email: \href{mailto:kannapinn@cps.tu-darmstadt.de}{kannapinn@cps.tu-darmstadt.de}}

\usepackage{amssymb}

\usepackage{lineno}

\usepackage{sectsty}
\subsectionfont{\normalfont\itshape}

\DeclareFieldFormat{citehyperref}{%
  \DeclareFieldAlias{bibhyperref}{noformat}
  \bibhyperref{#1}}

\DeclareFieldFormat{textcitehyperref}{%
  \DeclareFieldAlias{bibhyperref}{noformat}
  \bibhyperref{%
    #1%
    \ifbool{cbx:parens}
      {\bibcloseparen\global\boolfalse{cbx:parens}}
      {}}}

\savebibmacro{cite}
\savebibmacro{textcite}

\renewbibmacro*{cite}{%
  \printtext[citehyperref]{%
    \restorebibmacro{cite}%
    \usebibmacro{cite}}}

\renewbibmacro*{textcite}{%
  \ifboolexpr{
    ( not test {\iffieldundef{prenote}} and
      test {\ifnumequal{\value{citecount}}{1}} )
    or
    ( not test {\iffieldundef{postnote}} and
      test {\ifnumequal{\value{citecount}}{\value{citetotal}}} )
  }
    {\DeclareFieldAlias{textcitehyperref}{noformat}}
    {}%
  \printtext[textcitehyperref]{%
    \restorebibmacro{textcite}%
    \usebibmacro{textcite}}}

\begin{document}

\maketitle

\par\noindent\rule{\textwidth}{0.4pt}           
\begin{abstract} \noindent 
\vspace*{-1.6mm}
\paragraph*{Purpose}
\textcolor{black}{Simulation-based digital twins represent an effort to provide high-accuracy real-time insights into operational physical processes. However, the computation time of many multi-physical simulation models is far from real-time. It might even exceed sensible time frames to produce sufficient data for training data-driven reduced-order models. This study presents TwinLab, a framework for data-efficient, yet accurate training of neural-ODE type reduced-order models with only two data sets.} 

\paragraph*{Design/methodology/approach}
\textcolor{black}{Correlations between test errors of reduced-order models and distinct features of corresponding training data are investigated. Having found the single best data sets for training, a second data set is sought with the help of similarity and error measures to enrich the training process effectively.}

\paragraph*{Findings}

\textcolor{black}{Adding a suitable second training data set in the training process reduces the test error by up to 49\% compared to the best base reduced-order model trained only with one data set.
Such a second training data set should at least yield a good reduced-order model on its own and exhibit higher levels of dissimilarity to the base training data set regarding the respective excitation signal. Moreover, the base reduced-order model should have elevated test errors on the second data set. 
The relative error of the time series ranges from 0.18\% to 0.49\%. Prediction speed-ups of up to a factor of 36,000 are observed.} 

\paragraph*{Originality} 

\textcolor{black}{The proposed computational framework facilitates the automated, data-efficient extraction of non-intrusive reduced-order models for digital twins from existing simulation models, independent of the simulation software.} 

\end{abstract}

\par\noindent\rule{\textwidth}{0.4pt}\vspace*{2pt}
{\small
Accepted version of the revised manuscript published in \emph{Engineering Computations}. \\
Date of acceptance: May 23, 2024. DOI: \href{https://doi.org/10.1108/EC-11-2023-0855}{10.1108/EC-11-2023-0855}. 
License: \href{https://creativecommons.org/licenses/by-nc/4.0/legalcode}{CC BY-NC 4.0}
}
\vspace*{-1.6mm}
\par\noindent\rule{\textwidth}{0.4pt}


\widowpenalty10000
\clubpenalty10000

\renewcommand{\appendixname}{Appendix}

\newlength\fheight
\newlength\fwidth
\newlength\fheightthree
\newlength\fwidththree
\newlength\fheighttwo
\newlength\fwidthtwo
\newlength\figureheightbig
\newlength\figurewidthbig

\setlength\fheightthree{0.17\textheight}
\setlength\fwidththree{0.345\textwidth}

\setlength\fheighttwo{0.2\textheight}
\setlength\fwidthtwo{0.42\textwidth} 


\newpage
\section{Introduction}



Digital twins that are derived from multi-physical simulation data have the potential to drive a transformative shift towards more autonomous processes. 
To date, simulation and data science technologies are employed by experts in the design phase, primarily in the context of \emph{what-if} simulations to design and optimize products or processes. 
Digital twins represent an initiative of integrating insights from data and simulations into operational processes, empowering process control algorithms to perform informed, autonomous decision-making~\autocite{dt_Niederer2021,dt_Verboven2020,dt_Rasheed2020} and correcting the current operational conditions~\autocite{dt_moyaDigitalTwinsThat2022}.
Since its inception a decade ago by Grieves and Vickers~\autocite{dt_Grieves2017}, substantial efforts have been invested in shaping and standardizing the digital twin concept~\autocite{dt_stark2019,dt_AIAA2020,dt_ISO23247}, also establishing it as a dedicated research domain~\autocite{dt_tao2019,dt_Niederer2021}. 
In essence, a digital twin can be described as a virtual repository of information that mirrors its physical counterpart with the highest fidelity~\autocite{dt_Glaessgen2012,dt_Grieves2017,dt_Rasheed2020}. This digital-physical symbiosis thrives on real-time, bidirectional data exchange, spanning the entire lifecycle of a process or product~\autocite{dt_Niederer2021,dt_Rasheed2020,dt_AIAA2020,dt_Niederer2021,dt_Lu2020}.
Historically, approximately \qty{85}{\percent} of digital twin research has gravitated towards product life-cycle management. 
As the second largest area, \qty{11}{\percent} focused on factory or production planning~\autocite{dt_Lu2020}.
Nevertheless, digital twins have the potential to also serve as catalysts for the development of autonomous processes, ushering in a novel dimension in digital twin research and technology~\autocite{dt_Rosen2015}.

\textcite{dt_Niederer2021} assert the necessity of developing innovative mathematical, numerical, and computational methodologies to effectively implement digital twins on a large scale. 
As a crucial step in this direction, the present work introduces a methodology for deriving digital twins from multi-physical simulation models.
To ensure the faithful replication of their physical counterparts, digital twins should ideally rely on physics-based simulation models, given the advancements in computational engineering regarding accuracy and efficiency.
First-principle models, rooted in a deep understanding of cause-and-effect relationships, enable precise process control.
Integrating highly accurate simulation models with data-driven reduced-order modeling may be encapsulated in the term \enquote{physics-based, data-driven digital twin}~\autocite{diss_phd23}.

Requiring digital twins to replicate their physical counterparts in real time poses an enormous challenge.
Despite the significant growth in computing power, real-time simulation of large-scale industrial problems will remain unattainable in the coming years, and possibly even decades. 
Recognizing this dilemma since the 1990s, scientific computing and computational engineering researchers have identified reduced-order models (ROMs), also known as surrogates, as a promising solution~\autocite{rom_bennerMOR1}.
\textit{Data-driven} ROMs exclusively rely on the output data of the simulation model. These ROMs align well with the practicalities of model development within a diverse software landscape of open-source, commercial, or custom codes. 
Access to solvers in commercial simulation software is often limited, and even when the source code is available, intrusive ROMs may be considered excessively time-consuming~\autocite{rom_pehersdorfer2016}.
 \textcolor{black}{Still, certain machine-learning ROM methods demand extensive training data. Here, we face a novel dilemma of attempting online prediction speed increases with ROMs at the price of long offline waiting times for data. When employing time-consuming simulations, data production might exceed sensible time frames, especially for industrial product development cycles.}
This study introduces an approach to carefully selecting very few training data sets. 
An efficient design of experiments is proposed to guide the selection of two training data sets, ensuring minimal test error for ROMs when applied to representative test data.

\textcite{dt_Niederer2021} emphasized that the creation and dissemination of digital twins require open-source platforms to enable large-scale implementation.
\textcolor{black}{Some methods of the here-presented open-source software \emph{TwinLab} were developed as part of earlier research work~\autocite{diss_phd23,diss_twinlab,diss_ifset22}}.
TwinLab interconnects simulation models, training data selection, and control techniques within a unified framework, offering interfaces for two major commercial simulation software packages\textcolor{black}{: COMSOL Multiphysics and ANSYS Fluent.}
 \textcolor{black}{The current study concisely summarizes the reduced-order modeling procedure of TwinLab and extends this concept with the selection of additional data to enhance ROM training and further reduce test error.
TwinLab's functionality is demonstrated with a case study on generating digital twins for thermal food processing.}
The outline of this article is as follows. 
This section defines the essential attributes of digital twins and outlines the prerequisites for the ROM generation process. 
\autoref{sec:relatedwork} presents related works and previous research. Besides, it gives background information on the employed ROM method of \emph{neural ODE}-type.
\autoref{sec:methods} introduces the framework for  \textcolor{black}{reduced-order model} derivation from simulation data and briefly describes the use case. 
Results on training data selection are presented in \autoref{sec:results}, whereas a final conclusion and future steps are given in \autoref{sec:conclusion}.

\section{Related work} \label{sec:relatedwork}


\subsection{Training of transient reduced-order models} 

As the introduction highlights, the digital twin methodology must be accompanied by reduced-order modeling approaches capable of replicating multi-physical problems within a reasonable simulation time and with lean computational cost.
Given the diversity in how the problem is tackled across different disciplines, various perspectives exist on ROMs. A common classification criterion is the intrusiveness of the ROM generation method~\autocite{rom_bennerMOR1}. In this context, an \emph{intrusive} approach involves accessing and modifying the underlying set of PDEs. Conversely, methods that refrain from modifying the underlying equations are labeled as \emph{non-intrusive}~\autocite{rom_bennerMOR1,rom_pehersdorfer2016}.
In computational engineering, PDE-centric, intrusive approaches utilize the formulated PDE of the physical model as a starting point. 
After numerical discretization in space and time, the resulting equation is projected onto a reduced-order space, aiming for an accelerated solution to the problem~\autocite{rom_bennerMOR1}. 
However, commercial simulation software companies safeguard their solution algorithms as intellectual property, denying users root-level access to these algorithms. 
Consequently, intrusive ROM methods are not a feasible option. 
Therefore, this work \textcolor{black}{resorts} to non-intrusive ROM methods to maintain universality across employed simulation software.

Within the discipline of machine learning, time series \textcolor{black}{can be} replicated with recurrent neural networks (RNNs)~\autocite{rom_goodfellow16DL}.
Neural networks are known to be highly flexible, universal, nonlinear function approximators~\autocite{rom_goodfellow16DL}.
RNNs, more specifically, long short-term memory neural networks (LSTMs), are characterized by recursive calls of a neural network with a discrete temporal delay to invoke the progress of the variables in time. 
Through the back-propagation through time algorithm, the neural network is trained to replicate input-to-output relations over time.
Exemplary implementations of RNNs and discussions on overcoming difficulties with learning long-term dependencies can be found in~\autocite{rom_hochreiter2001}. 

Recently, there has been a growing interest in hybrid approaches that meld the concept of identifying system dynamics with machine-learning techniques. 
\textcite{rom_dupont2019ANODE} elucidate the resemblance between deep feed-forward neural networks, specifically residual networks~\autocite{rom_he2016resnet}, and differential equations. 
The mapping of a hidden state $\vect{h}_{t} \in \mathbb{R}^d$ at layer $t$ to its next layer is expressed as
\begin{align}
\vect{h}_{t+1} = \vect{h}_{t} + \vect{f}_{t} (\vect{h}_{t})\,,
\end{align}
where $\vect{f}_{t}: \mathbb{R}^d \rightarrow \mathbb{R}^d$ is a differentiable function projecting from one hidden state to the next. 
By forming a difference quotient and taking the limit of an imaginary time step, the similarity to an ODE system becomes evident:
\begin{align}
\lim _{\Delta t \rightarrow 0} \frac{\mathbf{h}_{t+\Delta t}-\mathbf{h}_t}{\Delta t}=\frac{\diff\mathbf{h}(t)}{\diff t}=\mathbf{f}(\mathbf{h}(t), t) \,.
\end{align}
An input is transformed to the output by solving an ODE over multiple time steps, with a feed-forward neural network representing the right-hand side operator $\vect{f}$ of the ODE. 
This approach is termed \emph{neural ODE}. 
Recent publications suggest that neural ODEs outperform RNNs or tree-based algorithms, as evidenced in \autocite{rom_Lu21node-vs-lstm-pharma} for predicting pharmacokinetics or in \autocite{rom_pepe22node-vs-lstm-battery} for predicting the remaining state of health of batteries. 
The success of neural ODEs is attributed to their ability to learn underlying dynamics rather than merely input-to-output relations~\autocite{rom_pepe22node-vs-lstm-battery}.

The software package \emph{ANSYS Dynamic ROM Builder} \autocite{rom_DynROMPatent,rom_twinbuilder_2020}, referred to as \emph{DynROM} hereafter, offers a non-intrusive, nonlinear, transient ROM using a method akin to neural ODEs. 
Given the promising results observed in preliminary testing and recognizing the need for in-depth investigations of the method in the literature, the decision was made to incorporate DynROM in this work.

\subsection{Design of experiments for data-driven ROMs} 

System identification is a type of reduced-order modeling that relies on tailored excitation signals to extract valuable information from physical models during either virtual or real experiments. These excitation signals enable the determination of the transient behavior of the physical system based on the recorded output data~\autocite{rom_Nelles2020}. There are a few notable works that cover the design of excitation signals for this purpose~\autocite{rom_Nelles2020,sig_Gringard2016,sig_Talis2021,sig_Heinz2017,sig_Heinz2018}. However, most research primarily guides designing experiments focused on nonlinear auto-regressive models with exogenous inputs (NARX), RNNs or proper orthogonal decomposition (POD) ROMs~\autocite{rom_guenotAdaptiveSamplingStrategies2013}.
No published work has been found that specifically addresses the generation and selection of training data for data-driven reduced-order modeling in the context of food models. 
In general, the selection of data sets for training reduced-order models of the \emph{neural ODE}-type to achieve low test errors remains largely unreported. 
Since the introduction of the ROM method in 2018, some works have employed this approach~\autocite{rom_Boscaglia2021,rom_calka2021,rom_kim2022dynromwind}. 
However, these works have not significantly focused on selecting training data.
To address these gaps, previous research~\autocite{diss_ifset22} proposed an efficient design of experiments using a single data set. 
Moreover, it highlighted that the conventional suggestion to uniformly cover the input space of models~\autocite{rom_Nelles2020} or their output space~\autocite{sig_Talis2021} to ensure effective training of reduced-order models does not apply to the specific food model presented. 
An input and output space coverage measure for the food model was implemented, and it was observed that high coverages did not correlate with low test errors~\autocite{diss_ifset22}.

\subsection{Digital twins and reduced-order modeling in food science}
Recent review articles have underscored the potential of digital twins in food science and technology~\autocite{dt_Verboven2020,dt_Defraeye2021coming,dt_henrichs2022}. 
Notably, \textcite{dt_henrichs2022} focused their review on the food value chain and shop floor production planning.
Their assessment shows few studies have delved into process autonomy within this domain.
Out of 84 pieces of research, only eight have centered on digital twin-enabled process autonomy, and a mere two were peer-reviewed studies.
In the post-harvest sector, several studies view digital twins as simulation models for processing temperature profiles of physical counterparts that have been pre-recorded~\autocite{dt_defraeye2019,dt_tagliavini2019,dt_shoji2022,dt_shrivastava2023twinpackage}. 
Concepts for establishing bi-directional linkages, interactive decision-making, and comprehensive product life cycle mirroring still need to be explored. 
The demand for real-time simulations, as recently envisioned~\autocite{dt_prawiranto2021,dt_Defraeye2021coming}, is becoming increasingly apparent.
For instance, in \autocite{dt_prawiranto2021}, the computational hardware, consisting of an 8-core Intel Core i7 processor with 32 GB of RAM, required  \SI{20}{hours} of simulation time to predict the solar drying of just one-eighth of a single apple ring within a non-conjugate simulation.
The achieved speed-ups, roughly three times faster than real-time, are arguably insufficient for supporting online decision-making with advanced control algorithms. 
Moreover, considering the computational cost, it would be infeasible to predict multiple scenarios, even for a single piece of fruit. 
This underscores the compelling need for \textcolor{black}{data-efficient} reduced-order modeling \textcolor{black}{to create digital twins}.
Precisely, data-driven reduced-order modeling, \textcolor{black}{irrespective of the simulation software}, constitutes a pivotal element.
While developing surrogate models is a well-established practice, the realm of food science notably lacks nonlinear, transient reduced-order models.
There exist exceptions, such as the works by \textcite{rom_Rivas2013} and \textcite{mpc_Alonso2013}, though they do not explicitly reference digital twins, they encompass various elements characteristic of digital twin systems: a physics-based simulation model, a reduced-order model, and a control algorithm.
A limitation of these works lies in the equation-invasive nature of the POD technique they employ and the necessity to address nonlinear terms~\autocite{rom_brunton22databook}. 
A substantial proportion of simulation models in food science are developed within commercial software environments, which often lack root-level access to equations for custom modifications. 
Consequently, the POD approach is not suitable for applications using commercial software.
In a more recent study by~\autocite{mpc_Alonso2021}, the authors noted that the solution times of their approach must be significantly reduced to achieve real-time optimal control.

In the context of food science, very few studies explore reduced-order models using machine-learning approaches to predict transient model behavior. 
For example, \textcite{rom_broyart2003rnn} trained RNNs to predict macroscopic quantities during biscuit baking based on experimental data.
\textcite{rom_Isleroglu2020predictioncookieANN} predicted the browning index of cookies using similar methods.
\textcite{rom_khan2022} recently conducted a comprehensive review of machine-learning-based modeling in food processing. 
Of the 31 studies presented, only two addressed transient reduced-order models~\autocite{mpc_Huang1998,mpc_Li2016}.
The remaining studies predominantly focus on steady-state machine-learning models in food science.
These models utilize experimental or simulation data to train feed-forward neural networks. Still, due to the absence of dynamics, they cannot influence process parameters during the surrogate model's execution.

\section{Methods} \label{sec:methods}


\subsection{\textcolor{black}{The digital twin use case of autonomous thermal food processing}} \label{sec:frameworkcooking}

Imagine a convection oven that is to carry out autonomous thermal food processing.
Evaluating food quality measures like core temperature, moisture content, or texture poses a challenge for the cooking appliance without additional equipment. 
Here, simple measurements of the current oven temperature can serve as the initial condition for predictions of the digital twin. 
A control algorithm can employ the control vector parameterization approach to adjust the amplitudes of a hypothetical oven temperature trajectory systematically. 
\textcolor{black}{With faster-than-real-time solution times, the underlying ROM swiftly predicts the food's temperature trajectories of those multiple future scenarios.}
This enables the algorithm to understand the model's sensitivities concerning a target function that includes desired cooking objectives. 
The optimization problem is solved at discrete points during operation to determine the optimal trajectory of oven temperatures to achieve the objectives. 
Repetitive comparisons between the actual oven temperature and the planned trajectory help mitigate the model-process mismatch by recalculating the residual trajectory.
This capability enables the device to plan its oven temperature trajectory to meet specific user requirements, such as achieving the desired moisture content, safe core temperatures, or texture at the end of the cooking process to a particular point in time.

A soft-matter model for chicken meat~\autocite{cm_rabeler_mod2018,cm_feyissa_3D2013} has been implemented for this case study. 
This model consists of two coupled transport equations for heat transfer (temperature $T$) and moisture transport (moisture content on a dry basis $M_\text{db}$). 
The convective heat influx from the oven is introduced using a heat transfer coefficient $\alpha$ within a mixed boundary condition \textcolor{black}{in conjunction with the prescribed $T_\text{oven}$}. 
For more extensive implementation details, sensitivities of model parameters, and validation tests, refer to previous research~\autocite{diss_ifset22,diss_phd23}.

This case study focuses on replicating temperature probes at specific points of the full-order model, as illustrated in \autoref{fig:fw-overview}. 
By examining point data reduced-order models, this study establishes a condensed and minimalistic setup that allows for a thorough investigation of the effects of training data selection.
In line with best practices~\autocite{rom_twinbuilder_2020}, only temperatures are chosen for training to eliminate the influence of different physics and scales within a single reduced-order model. 
Specifically, core temperatures $T_\text{A}$ and surface temperature $T_\text{B}$ are selected, representing the potential temperature trajectories within the model. 
Core temperatures are commonly monitored to assess the degree of cooking. 
For example, the U.S. Food and Drug Administration mandates holding times of at least one second at temperatures above $\SI{74}{\degreeCelsius}$~\autocite{mot_foodcode2017}. 
Surface temperatures can be significant for assessing the temperature-dependent browning of the food.

\subsection{\textcolor{black}{The TwinLab framework for reduced-order model derivation}} \label{sec:framework}

\begin{figure}[t]
     \centering
     \includegraphics[width=\textwidth]{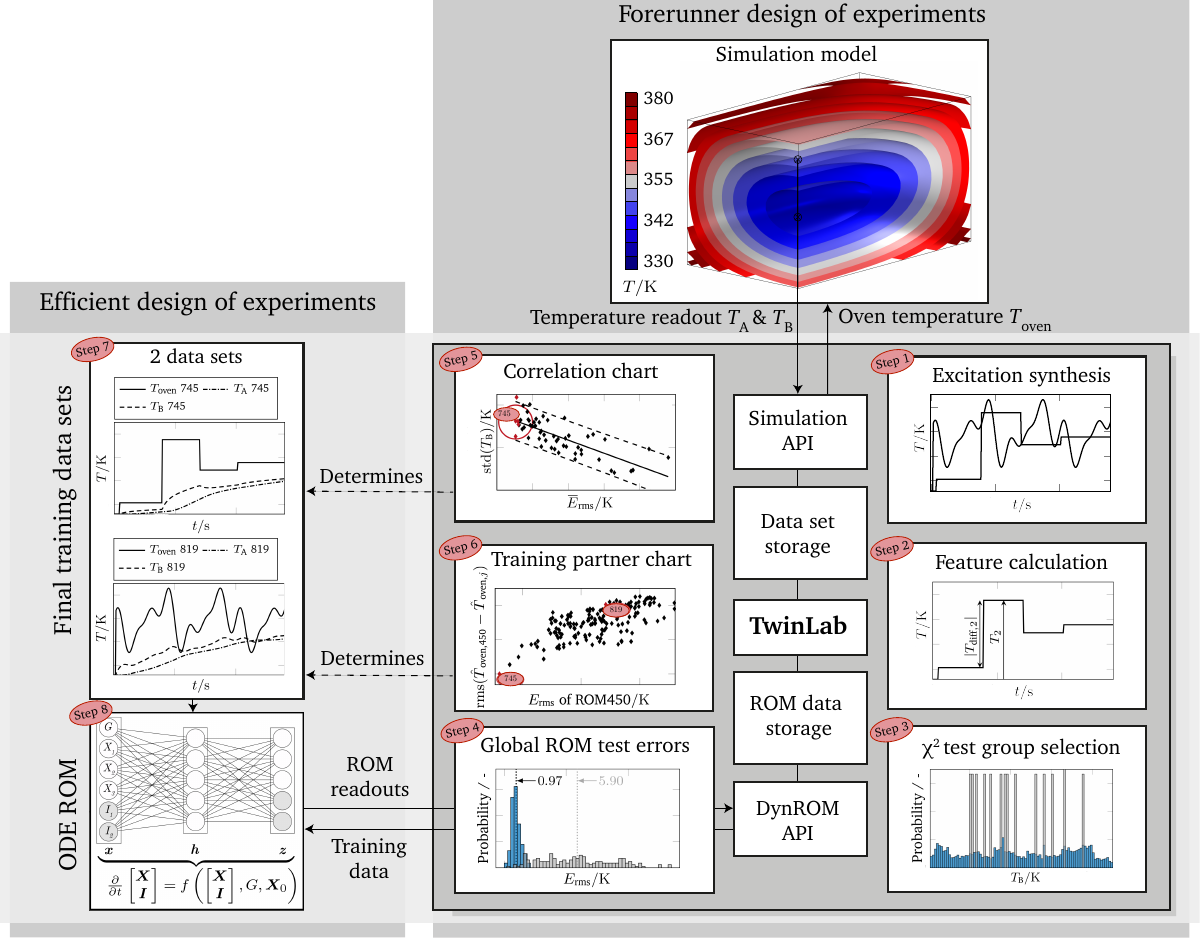}
    \caption{The proposed physics-based, data-driven digital twin framework. Source: Created by author.}
    \label{fig:fw-overview}
 \end{figure}

The central building block of the framework forms the data-driven derivation of a reduced-order model from the multi-physical simulation data following a procedure illustrated in \autoref{fig:fw-overview}: 
\begin{outline}
\1 Step 1: Synthesized excitation signals are employed to vary the oven temperature in the simulations. 
\2 Possible excitations include amplitude-modulated pseudo-random binary sequences (APRBS) or multi-sines, both suitable for nonlinear system identification~\autocite{rom_Nelles2020}. Previous research has demonstrated the suitability of APRBS signals~\autocite{diss_ifset22,diss_phd23}. 
\2 A basis of simulation data sets is automatically generated with different excitation signals and is stored in the \emph{data set storage}. 
\2 Each data set comprises a unique combination of excitation signal and simulation output data, as visualized in the left column of \autoref{fig:fw-overview} 
\1 Step 2: Feature calculation of all data sets.
\2  Several data set features are derived in the hope that at least one feature will correlate with low ROM test errors. For a complete list of features, refer to \autocite{diss_ifset22,diss_phd23}.
\2  Examples of possible features include the standard deviation and mean value of $T_\text{oven},\,T_\text{A}$ and $T_\text{B}$ or the signed mean of the delta jumps of an APRBS signal $\overbar{T}_{\text{diff},j}  \defeq  \operatorname{mean}([T_{\text{diff},2},T_{\text{diff},3},T_{\text{diff},4}])$.

\1 Step 3:  Representative data sets are selected with a $\chi^2$ test to serve exclusively for ROM testing on unseen data.

\1 Step 4: One data set at a time is used to train a corresponding ROM. Consecutively, \textcolor{black}{average} test errors are calculated for each ROM.

\1 Step 5: The Pearson correlation matrix reveals correlations between training data set features and the corresponding ROM's prediction errors. 
\2 These identified correlations help to determine the best single data set. Additionally, the knowledge of decisive features that correlate with low test errors for a particular ROM enables the synthesis of even better data sets. \autoref{sec:corrs} provides further details on this step. 

\1 Step 6: The training partner chart aids with selecting appropriate combinations of two data sets to improve ROM training and further reduce prediction errors. \autoref{sec:signalcombo} focusses on this step. 

\1 Step 7: Two final data sets can be determined with the knowledge of correlations and promising data set combinations.

\1 Step 8: From this data, the final ROM is trained and exported in the functional mockup unit (FMU) format~\autocite{dt_fmu}.
\end{outline}

\textcolor{black}{
The computational time required for simulations of realistic-sized multi-physical problems can extend over weeks or even months when executed on modern cluster PCs~\autocite{diss_phd23}. 
Consequently, TwinLab uses a forerunner concept: rather than conducting the time-intensive correlation search using the final full-order model, a priori correlation sampling is undertaken on a reduced-size simulation model, such as the one outlined in this study. 
Typically, such preliminary models emerge as a natural outcome during the development of the simulation model.
In computational engineering, it is uncommon to directly formulate a new simulation model encompassing all physical couplings and integrating detailed geometry of the problem employing a fine mesh. 
 Instead, the validation of individual components of the model occurs incrementally. 
Following the correlation search, only final full-order model simulations of the two most promising training data candidates are necessary.
In a prior investigation~\autocite{diss_phd23}, a trend in the test error of reduced-order models can be established between the preliminary and full-order models, provided that both models feature identical physical models and operational conditions to ensure comparability.
}


\subsection{Model order reduction with neural-ODE-type ROMs}\label{sec:rommethod}


\textcolor{black}{The procedure to infer a ROM for the thermal food processing use case, as illustrated in \autoref{fig:fw-overview}, can be outlined as follows:}
The thermal food processing model, referred to as the full-order model, is automatically simulated through command line calls within TwinLab. 
The discretized oven temperatures $\vect{\disphat{G}} = [T_{\text{oven},1}, \dots, T_{\text{oven},N}]$ serve as external excitations for the full-order model over time, where $N$ represents the total number of time steps. 
Virtual probes read temperatures, such as \textcolor{black}{a core and surface temperature} $T_{\text{A},k}$ and $T_{\text{B},k}$ at discrete points in space and time, storing them in an array $\vect{\disphat[1.0mu]{Y}} \in \mathbb{R}^{n \times N}$ ($n=2$ in this case).
Subsequently, the ROM is excited similarly by the oven temperatures $\vect{\disphat{G}}$ at its input. 
During a training phase, the ROM's parameters are optimized to ensure that the discretized ROM output $\vect{\disphat{X}}$ replicates the full-order model output: $\vect{\disphat[1.0mu]{Y}} \approx \vect{\disphat{X}}$. 
The input $G=G(t)$ is mapped to the output $\vect{X} = \vect{X}(t)$ by solving the ODE:
\begin{align}\label{eq:DynROMODE}
\frac{\partial}{\partial t} \begin{bmatrix} \vect{X} \\ \vect{I} \end{bmatrix} &= \vect{f} \left( \begin{bmatrix} \vect{X}\\ \vect{I}\end{bmatrix}, G,\vect{X}_0 \right),\\
    \vect{X}(t=0) &= \vect{X}_0 \,,
\end{align}
where the state vector $\vect{X} \in \mathbb{R}^{n}$ is extended by a vector of additional variables $\vect{I} \in \mathbb{R}^{i}$ --- introduced later --- and $\vect{X}_0 \in \mathbb{R}^{n+i}$ contains the initial conditions for both.

A three-layer feed-forward neural network represents the right-hand side operator $\vect{f}$. 
The neural network comprises an input layer $\vect{x} \in \mathbb{R}^{n+i+1}$, a hidden layer $\vect{h} \in \mathbb{R}^{n+i}$, and an output layer $\vect{z} \in \mathbb{R}^{n+i}$, as depicted in the bottom left of \autoref{fig:fw-overview}. 
Linear transformations of the input layer $\vect{x}$ with weight matrices $\vect{W}_1$ and the addition of a bias vector $\vect{b}_1$ constitute the fundamental operations to compute the values of the neurons in the hidden layer $\vect{h}$. 
Applying a sigmoid activation function $\mathscr{S}$ introduces nonlinearity to the relationship between the layers~\autocite{rom_goodfellow16DL}. 
This process is repeated for the output layer, involving different weights and biases, leading to the computation from the input to the output layer:
\begin{align}
\vect{h} &= \mathscr{S}(\vect{W}_1 \vect{x} + \vect{b}_1) \,, \\
\vect{z} &= \mathscr{S}(\vect{W}_2 \vect{h} + \vect{b}_2) \,.
\end{align}
To establish the desired relationship between the input and output layers, training data with known input $\vect{\disphat{G}}$ and output $\vect{\disphat[1.0mu]{Y}}$ is presented to the neural network. 
This process, known as supervised learning, involves implementing fourth-order Runge--Kutta schemes in DynROM to numerically integrate the ODE system over time. 
The loss function for evaluating the neural network's training error is the mean squared error:
\begin{align}
    E_\text{mse} =  \frac{1}{n} \sum_{j=1}^n \left(\frac{1}{N} \sum_{k=1}^{N} (\disphat{X}_{jk} - \disphat[1.0mu]{Y}_{jk})^2 \right)\,,
\end{align}
that is averaged across all learning scenarios.
To minimize this loss function, gradient descent algorithms are employed in conjunction with the back-propagation algorithm \autocite{rom_brunton22databook}. 
This optimization process aims to determine optimal values for the weights and biases in all layers of the neural network, ensuring that the discrete output $\vect{\disphat{X}}$ obtained from numerically integrating \autoref{eq:DynROMODE} closely replicates the output of the full-order model $\vect{\disphat[1.0mu]{Y}}$.

ODEs inherently feature a vector field on their right-hand side, prohibiting trajectories from different initial conditions to intersect. 
\textcite{rom_dupont2019ANODE} demonstrated that neural ODEs struggle to learn crossing paths from input to output without special augmentation. 
This limitation was addressed by introducing additional free variables $\vect{I}$ to $\vect{X}$ \autocite{rom_dupont2019ANODE}, resulting in what was termed \emph{augmented neural ODEs}. 
Augmentation involves adding extra neurons to the layers, enabling trajectories to be lifted into additional dimensions where they no longer need to cross. 
The term \emph{complexity} $i$ is used in this work to denote the number of added free variables.
Similar to the augmentation strategy of neural ODEs, DynROM incorporates $i$ free variables to the state vector, as long as it contributes to reducing the training error \autocite{rom_twinbuilder_2020}. 
Notably, the method exhibits a unique characteristic of requiring minimal training data, such as one or two simulations over time, each containing a moderate number of time steps (e.g., $N=280$ in the studies of \autoref{sec:results}).

\section{Case study on training data selection} \label{sec:results}


\subsection{Correlation-based training data selection  \pages{5}}\label{sec:corrs}

This section summarizes prior research \autocite{diss_ifset22,diss_phd23}, laying the foundation for the case study on combinations of training data sets presented in this work. 
The framework's procedure, as introduced in \autoref{sec:framework}, is followed until step 5 to identify correlations between global error measures and training data features.
Fifty-five amplitude-modulated pseudo-random binary sequence (APRBS) excitation signals are employed in the full-order simulation model to vary $T_\text{oven}$ and generate readouts as defined in \autoref{sec:rommethod}. 
Each data set is uniquely identified with an alphanumeric identifier based on consecutive numbering in TwinLab.
After step 3, 15 data sets (referred to as AP15 hereafter) are automatically selected for testing. 
Subsequently, all data sets are individually used to train a \emph{1-data-set ROM.} 
The median of all training errors, i.e., the root-mean-square of the difference between ROM and the corresponding training data set output, is $\SI{0.22}{K}$, with a standard deviation of $\SI{0.14}{K}$.
Six global error measures are derived from testing the ROMs, including the average over all test data (indicated by $\overline{(\cdot)}$) for the root-mean-square error $\rmseq$, the mean absolute percentage error $\mapeq$, the maximum error $\maxeq$, the median error $\medeq$, the interquartile range $\iqreq$, and the coefficient of determination R$^2$ (see the first column in \autoref{tab:Corr-Show-a4-Eval-APRBSEqual15} for the corresponding variables).

\renewrobustcmd{\bfseries}{\fontseries{b}\selectfont}
\newrobustcmd{\B}{\bfseries}

%
\begin{table*}[b]
\small
\addtolength{\tabcolsep}{-2.1pt}
\sisetup{detect-weight,     
         mode=text,         
         table-format=$-$0.4, 
         add-integer-zero=false,
         table-space-text-post={*} 
         }
\centering
 \caption{Pearson correlation matrix between error measures (rows) and training data features (columns) for APRBS 1-data-set ROMs (testing on AP15). $\operatorname{std}(T_\text{B})$ shows the best a posteriori correlation whereas $\overbar{T}_{\text{diff},j}$ is a potential a priori data set feature. Source: Created by author.} 
 \vspace{1em}
 \begin{tabular}{lrrrrrrrrrrrrrr} \toprule 
Measure & $\overbar{T}_i$ & $\overbar{T}_{\text{diff},i}$ & $\overbar{T}_{\text{diff},j}$ & $| \overbar{T}_{\text{diff},i} |$ & $\overbar{T}_\text{oven}$ & $\operatorname{std}(T_\text{oven})$ & $\operatorname{Cr}(T_\text{oven})$ &  $\operatorname{std}(T_\text{A})$  & $\operatorname{std}(T_\text{B})$ \\ 
\midrule 
$\rmseg / \unit{K}$    & 0.18            & $-$0.39                 & \B{$-$0.68}                & 0.40                    & 0.33                    & $-$0.12              & $-$0.13                  & $-$0.05             & \B$-$0.76           \\
$\mapeg / \unit{\%}$   & 0.20            & $-$0.38                 & \B$-$0.69                 & 0.43                    & 0.35                    & $-$0.11              & $-$0.14                    & $-$0.06             & \B$-$0.78             \\
$\maxeg / \unit{K}$   & 0.13            & $-$0.47                 & \B$-$0.69                 & 0.33                    & 0.25                    & $-$0.13              & $-$0.10                    & 0.03              & \B$-$0.70             \\
$\medeg / \unit{K}$  & 0.19            & $-$0.37                 & \B$-$0.72                 & 0.50                    & 0.35                    & $-$0.09              & $-$0.13                    & $-$0.11             &\B $-$0.83             \\
$\iqreg / \unit{K}$   & 0.31            & $-$0.07                 &\B $-$0.32                 & 0.23                    & 0.36                    & $-$0.02              & $-$0.07                              & $-$0.08             & \B$-$0.37             \\
$\overline{\text{R}^2} $      & $-$0.23           & 0.40                  & \B0.60                  & $-$0.29                   & $-$0.37                   & 0.13               & 0.14                         & $-$0.08             & \B0.61              \\
\bottomrule 
 \label{tab:Corr-Show-a4-Eval-APRBSEqual15} \end{tabular}\end{table*}

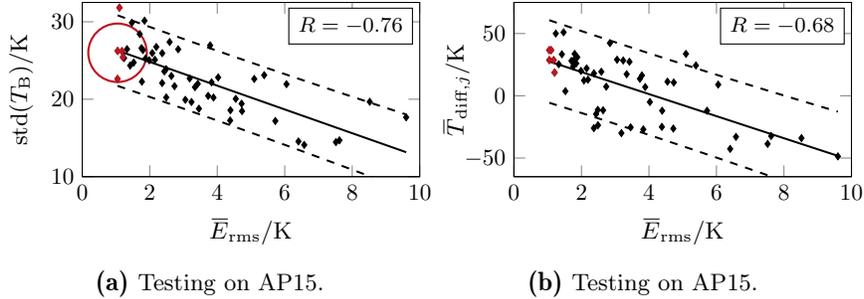
\begin{figure}[t]
\centering
\setlength\fheight{\fheightthree}
\setlength\fwidth{\fwidththree}
\begin{subfigure}[b]{0.32\textwidth}
\setlength\fheight{\fheightthree}
\setlength\fwidth{\fwidththree}
%
%
\definecolor{mycolor1}{rgb}{0.46600,0.67400,0.18800}%
\definecolor{mycolor2}{rgb}{0.00000,0.61176,0.85490}%
\definecolor{mycolor3}{rgb}{0.30100,0.74500,0.93300}%
\definecolor{mycolor4}{rgb}{0.96078,0.63922,0.00000}%
\definecolor{mycolor5}{rgb}{0.63500,0.07800,0.18400}%
\definecolor{mycolor6}{rgb}{0.00000,0.44700,0.74100}%
\definecolor{mycolor7}{rgb}{0.75294,0.75294,0.75294}%
\definecolor{mycolor8}{rgb}{0.85000,0.32500,0.09800}%
\definecolor{mycolor9}{rgb}{0.92900,0.69400,0.12500}%
\definecolor{mycolor10}{rgb}{0.65098,0.00000,0.51765}%
%
\begin{tikzpicture}
\begin{axis}[%
xmin=0.0,
xmax=10,
xlabel={$\rmseg / \unit{K}$},
ymin=10,
ymax=32.5,
ylabel={$\operatorname{std}(T_\text{B}) / \unit{K}$},
axis background/.style={fill=white},
title style={font=\bfseries},
legend style={at={(0.99,0.98)}, anchor=north east, legend cell align=left, align=left, draw=white!15!black},
width = \fwidth, 
 height = \fheight 
]

%

\addplot [color=black]
  table[]{../tiks/ROM-signal-evals/APRBSEqual15-RMSE-stdTb-Show-a4-2.tsv};
\addlegendentry{Fit APRBS}

\addplot [color=black, dashed]
  table[]{../tiks/ROM-signal-evals/APRBSEqual15-RMSE-stdTb-Show-a4-3.tsv};
  \addlegendentry{$\text{Fit p \textgreater{} 0.95}$}
  \addplot [color=black, dashed, forget plot]
  table[]{../tiks/ROM-signal-evals/APRBSEqual15-RMSE-stdTb-Show-a4-4.tsv};

%
%
%

\addplot [color=mycolor8, only marks, mark=diamond*, mark options={solid,mark size=1pt, fill=black, black}]
  table[]{../tiks/ROM-signal-evals/APRBSEqual15-RMSE-stdTb-Showsr5nl473f0005-a4-m1-m2-m3-m4-m5-m6t-m7-m8-m9-m10schroeder-9.tsv};
\addlegendentry{APRBS}

\addplot [color=tud10a, only marks, mark=diamond*, mark options={solid,mark size=1pt, fill=tud10a, tud10a}]
  table[]{../tiks/ROM-signal-evals/APRBSEqual15-RMSE-stdTb-Showsr5nl473f0005-a4-m1-m2-m3-m4-m5-m6t-m7-m8-m9-m10schroeder-10.tsv};
\addlegendentry{Best 5 APRBS}

\legend{}
\node[fill=white, below right, align=right, draw=black,anchor=north east, font=\scriptsize]
at (rel axis cs:0.98,0.97) {$R = -0.76$}; 

\node[draw,circle,thick, anchor=center, align=center, draw=tud10a,inner sep=3pt,font=\small,scale=2]
at (axis cs:1.04588031583495,26) {~};

\end{axis}

\begin{axis}[%
xmin=0,
xmax=1,
ymin=0,
ymax=1,
axis line style={only marks},
ticks=none,
axis x line*=bottom,
axis y line*=left,
legend style={legend cell align=left, align=left, draw=white!15!black},
width = \fwidth, 
 height = \fheight 
]

\end{axis}
\end{tikzpicture}%
 \caption{Testing on AP15.} \label{fig:KPIandErrorCorrelations-a}
 \end{subfigure} 
%
%
\begin{subfigure}[b]{0.32\textwidth}
\setlength\fheight{\fheightthree}
\setlength\fwidth{\fwidththree}
%
%
\definecolor{mycolor1}{rgb}{0.46600,0.67400,0.18800}%
\definecolor{mycolor2}{rgb}{0.30100,0.74500,0.93300}%
\definecolor{mycolor3}{rgb}{0.65098,0.00000,0.51765}%
\begin{tikzpicture}

\begin{axis}[%
xmin=0.0,
xmax=10,
xlabel={$\rmseg / \unit{K}$},
 y label style={yshift=-.79em},
ymin=-65,
ymax=75,
ylabel={$\overbar{T}_{\text{diff},j} / \unit{K}$},
axis background/.style={fill=white},
title style={font=\bfseries},
legend style={at={(0.97,0.03)}, anchor=south east, legend cell align=left, align=left, draw=white!15!black},
width = \fwidth, 
 height = \fheight 
]

\addplot [color=black]
  table[]{../tiks/ROM-signal-evals/APRBSEqual15-RMSE-mean-d-Ai-2-Show-a4-2.tsv};
\addlegendentry{Fit}

\addplot [color=black, dashed]
  table[]{../tiks/ROM-signal-evals/APRBSEqual15-RMSE-mean-d-Ai-2-Show-a4-3.tsv};
\addlegendentry{$\text{Fit p \textgreater{} 0.95}$}

\addplot [color=black, dashed, forget plot]
  table[]{../tiks/ROM-signal-evals/APRBSEqual15-RMSE-mean-d-Ai-2-Show-a4-4.tsv};
\addplot [color=mycolor1, only marks, mark=diamond*, mark options={solid,mark size=1pt, fill=black, black}]
  table[]{../tiks/ROM-signal-evals/APRBSEqual15-RMSE-mean-d-Ai-2-Show-a4-5.tsv};
\addlegendentry{APRBS}

\addplot [color=tud10a, only marks, mark=diamond*, mark options={solid,mark size=1pt, fill=tud10a, tud10a}]
  table[]{../tiks/ROM-signal-evals/APRBSEqual15-RMSE-mean-d-Ai-2-Show-a4-6.tsv};
\addlegendentry{Best 5 APRBS}

\legend{};
\node[fill=white, below right, align=right, draw=black,anchor=north east, font=\scriptsize]
at (rel axis cs:0.98,0.97) {$R = -0.68$}; 
\end{axis}

\begin{axis}[%
xmin=0,
xmax=1,
ymin=0,
ymax=1,
axis line style={only marks},
ticks=none,
axis x line*=bottom,
axis y line*=left,
legend style={legend cell align=left, align=left, draw=white!15!black},
width = \fwidth, 
 height = \fheight 
]

\end{axis}
\end{tikzpicture}%
 \caption{Testing on AP15.} \label{fig:KPIandErrorCorrelations-c}
\end{subfigure} 


%
 \caption{Correlations of signal features and $\rmseg$ of 1-data-set ROMs for APRBS (\protect\includegraphics[scale=0.4]{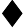}) and best 5 APRBS (\protect\includegraphics[scale=0.4]{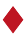}) training data sets. Solid and dashed lines are the linear regression curve and $p=0.95$ bounds. The red circle indicates the position of 1-data-set ROMs with low test errors. Source: Created by author.}
 \label{fig:KPIandErrorCorrelations}
\end{figure}

The Pearson correlation coefficient $R$ between training error and global test measures is 0.096, indicating hardly any correlation. 
Consequently, the training error does not significantly impact the study outcomes concerning the global error measures of the ROMs. 
This lack of correlation is relevant for maintaining a neutral study design. 
Moving forward, the Pearson correlation matrix will unveil correlations between training data features and global test error measures (as per step 5), detailed in \autoref{tab:Corr-Show-a4-Eval-APRBSEqual15}.
Numbers close to a magnitude of one imply a strong correlation, and these are highlighted in bold fonts in \autoref{tab:Corr-Show-a4-Eval-APRBSEqual15}. 
The correlation trends presented hold not only for the test data set AP15 but also for testing on additional data sets that were excited with APRBS and other signals, as shown in previous research~\autocite{diss_ifset22,diss_phd23}.
The most significant correlation is found for $\operatorname{std}(T_\text{B})$. 
A correlation coefficient value of $R=-0.76$ between $\operatorname{std}(T_\text{B})$ and $\rmseg$ indicates that high standard deviations in the surface temperatures of the full-order model correlate with low 1-data-set ROM test errors. 
For better visual interpretation, \autoref{fig:KPIandErrorCorrelations-a} depicts the correlation between $\operatorname{std}(T_\text{B})$ and $\rmseg$, where each marker symbolizes the evaluation of one 1-data-set ROM.
In the class of a priori training data features, $\overbar{T}_{\text{diff},j}$ also shows a good correlation to error measures, as seen in \autoref{fig:KPIandErrorCorrelations-c}. 
This feature reveals that APRBS signals with a globally ascending trend result in lower global ROM test errors. 
$\overbar{T}_{\text{diff},j}$ is the mean signed sum of delta jumps of an APRBS signal, where the first jump is excluded from the calculation. 
These correlations for $\operatorname{std}(T_\text{B})$ and $\overbar{T}_{\text{diff},j}$ illustrate how the physical particularities of a model influence the efficient design of experiments for data-driven reduced-order modeling through the search for correlations.
Achieving a high standard deviation in surface temperatures requires a more dynamic variation of oven temperatures. 
This underscores the need to incorporate the maximum operational conditions within a single training dataset. 
However, the low direct correlation of $T_\text{oven}$ with the global test errors ($R=-0.12$) highlights the intricate nature of synthesizing appropriate training data for nonlinear system identification.

To summarize, the identified correlations now offer insights into synthesizing or pre-selecting training data based on the proposition that $\operatorname{std}(T_\text{B})$ and $\overbar{T}_{\text{diff},j}$ should exhibit high values. 
The best five training signals for a 1-data-set ROM are highlighted in red in \autoref{fig:KPIandErrorCorrelations}. 
Additionally, the red circle indicates the location of promising training data for a 1-data-set ROM. 
Data set 745 emerges as the most effective training dataset for a 1-data-set ROM within the APRBS category and overall.

\subsection{Combination of training data sets} \label{sec:signalcombo}
\begin{figure*}[t] 
 \centering


     \begin{subfigure}[]{0.47\textwidth}
     \centering
   \setlength\fheight{0.22\textheight}
 \setlength\fwidth{0.95\textwidth}
\hspace{-1.5em}\input{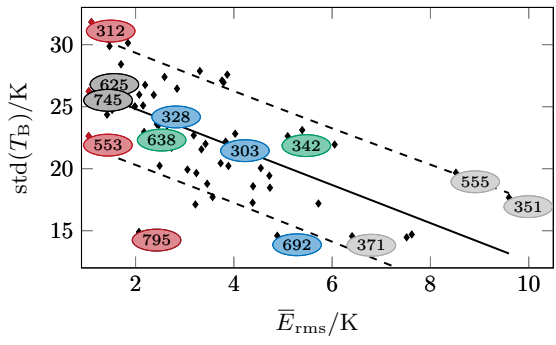}
      \caption{Test errors of all 1-data-set ROMs on AP15. }\label{fig:combo-a}
    \end{subfigure}
      \begin{subfigure}[]{0.49\textwidth}
  \centering
     \setlength\fheight{0.22\textheight}
 \setlength\fwidth{0.95\textwidth}
    \input{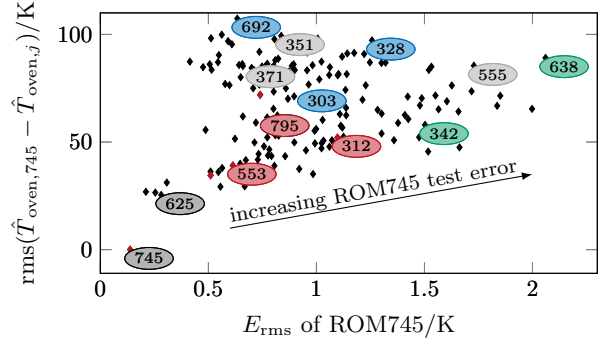}
    \caption{Signal similarity to 745 over $\rmseq$ of ROM745.}\vspace{0.8em}\label{fig:combo-b}
    \end{subfigure}
   \begin{subfigure}[]{0.965\textwidth}
  \centering
  \vspace{1em}
   \setlength\fwidth{\textwidth}
  \setlength\fheight{0.22\textheight}
    \input{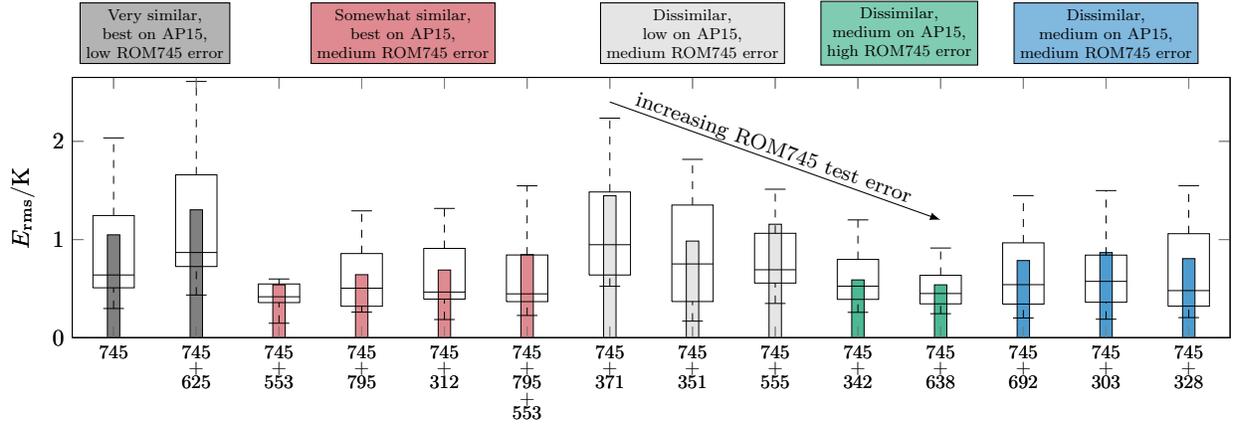}
    \caption{Boxplots and average test errors (represented by bars) of ROMs trained with different data set combinations are presented. The testing is conducted on AP15, and similar trends are observed for test data sets MS15 and sinAP15~\autocite{diss_phd23}. 
    }\label{fig:combo-c}
    \end{subfigure}
  \caption{Data sets performing well on AP15 (on the left side) are potential candidates for training partner selection. On the other hand, data sets sufficiently dissimilar to signal 745 are favorable as training partners. These observations highlight the importance of identifying data sets that perform well on the selected metrics while introducing diversity to enhance the overall ROM training accuracy. Source: Created by author.
} \label{fig:Combo-Comparison}
\end{figure*}

\begin{figure*}[t] 
\centering
\captionsetup[subfigure]{aboveskip=-1pt,belowskip=5pt}
\begin{subfigure}[]{\textwidth}
\centering
\setlength\fheight{0.15\textheight}
\setlength\fwidth{0.28\textwidth}
%
%
%
\begin{tikzpicture}
\pgfplotsset{
compat=1.11,
legend image code/.code={
\draw[mark repeat=2,mark phase=2]
plot coordinates {
(0cm,0cm)
(0.15cm,0cm)        
(0.3cm,0cm)         
};%
}
}
\begin{axis}[%
xmin=0,
xmax=1400,
xlabel={$t/\unit{s}$},
ymin=278,
ymax=475,
ylabel={$T / \unit{K}$},
axis background/.style={fill=white},
title style={font=\bfseries},
label style={font=\footnotesize},
legend style={at={(0.02,0.97)},anchor=north west, legend cell align=left, align=left, font=\scriptsize,inner sep=1.5pt},
width = \fwidth, 
 height = \fheight 
]
\addplot [color=black]
  table[]{../tiks/datasets/dspartners-1.tsv};

\addplot [color=black, densely dashdotted]
  table[]{../tiks/datasets/dspartners-2.tsv};

\addplot [color=black, densely dashed]
  table[]{../tiks/datasets/dspartners-3.tsv};
  
\node[fill=white, below right, align=right, draw=black,anchor=north east, font=\tiny]
at (rel axis cs:0.98,0.97) {$303$};

\end{axis}

\end{tikzpicture}%
%
%
%
\begin{tikzpicture}
\pgfplotsset{
compat=1.11,
legend image code/.code={
\draw[mark repeat=2,mark phase=2]
plot coordinates {
(0cm,0cm)
(0.15cm,0cm)        
(0.3cm,0cm)         
};%
}
}
\begin{axis}[%
xmin=0,
xmax=1400,
xlabel={$t/\unit{s}$},
ymin=278,
ymax=475,
axis background/.style={fill=white},
title style={font=\bfseries},
label style={font=\footnotesize},
legend style={at={(0.02,0.97)},anchor=north west, legend cell align=left, align=left, font=\scriptsize,inner sep=1.5pt},
width = \fwidth, 
 height = \fheight 
]
\addplot [color=black]
  table[]{../tiks/datasets/dspartners-4.tsv};

\addplot [color=black, densely dashdotted]
  table[]{../tiks/datasets/dspartners-5.tsv};

\addplot [color=black, densely dashed]
  table[]{../tiks/datasets/dspartners-6.tsv};

\node[fill=white, below right, align=right, draw=black,anchor=north east, font=\tiny]
at (rel axis cs:0.98,0.97) {$312$};
\end{axis}

\end{tikzpicture}%
%
%
%
\begin{tikzpicture}
\pgfplotsset{
compat=1.11,
legend image code/.code={
\draw[mark repeat=2,mark phase=2]
plot coordinates {
(0cm,0cm)
(0.15cm,0cm)        
(0.3cm,0cm)         
};%
}
}
\begin{axis}[%
xmin=0,
xmax=1400,
xlabel={$t/\unit{s}$},
ymin=278,
ymax=475,
axis background/.style={fill=white},
title style={font=\bfseries},
label style={font=\footnotesize},
legend style={at={(0.02,0.97)},anchor=north west, legend cell align=left, align=left, font=\scriptsize,inner sep=1.5pt},
width = \fwidth, 
 height = \fheight 
]
\addplot [color=black]
  table[]{../tiks/datasets/dspartners-7.tsv};

\addplot [color=black, densely dashdotted]
  table[]{../tiks/datasets/dspartners-8.tsv};

\addplot [color=black, densely dashed]
  table[]{../tiks/datasets/dspartners-9.tsv};

\node[fill=white, below right, align=right, draw=black,anchor=north east, font=\tiny]
at (rel axis cs:0.98,0.97) {$328$};
\end{axis}

\end{tikzpicture}%
%
%
%
\begin{tikzpicture}
\pgfplotsset{
compat=1.11,
legend image code/.code={
\draw[mark repeat=2,mark phase=2]
plot coordinates {
(0cm,0cm)
(0.15cm,0cm)        
(0.3cm,0cm)         
};%
}
}
\begin{axis}[%
xmin=0,
xmax=1400,
xlabel={$t/\unit{s}$},
ymin=278,
ymax=475,
axis background/.style={fill=white},
title style={font=\bfseries},
label style={font=\footnotesize},
legend style={at={(0.02,0.97)},anchor=north west, legend cell align=left, align=left, font=\scriptsize,inner sep=1.5pt},
width = \fwidth, 
 height = \fheight 
]
\addplot [color=black]
  table[]{../tiks/datasets/dspartners-10.tsv};

\addplot [color=black, densely dashdotted]
  table[]{../tiks/datasets/dspartners-11.tsv};

\addplot [color=black, densely dashed]
  table[]{../tiks/datasets/dspartners-12.tsv};

\node[fill=white, below right, align=right, draw=black,anchor=north east, font=\tiny]
at (rel axis cs:0.98,0.97) {$342$};
\end{axis}

\end{tikzpicture}%

%
%
%
\begin{tikzpicture}
\pgfplotsset{
compat=1.11,
legend image code/.code={
\draw[mark repeat=2,mark phase=2]
plot coordinates {
(0cm,0cm)
(0.15cm,0cm)        
(0.3cm,0cm)         
};%
}
}
\begin{axis}[%
xmin=0,
xmax=1400,
xlabel={$t/\unit{s}$},
ymin=278,
ymax=475,
ylabel={$T / \unit{K}$},
axis background/.style={fill=white},
title style={font=\bfseries},
label style={font=\footnotesize},
legend style={at={(0.02,0.97)},anchor=north west, legend cell align=left, align=left, font=\scriptsize,inner sep=1.5pt},
width = \fwidth, 
 height = \fheight 
]
\addplot [color=black]
  table[]{../tiks/datasets/dspartners-13.tsv};

\addplot [color=black, densely dashdotted]
  table[]{../tiks/datasets/dspartners-14.tsv};

\addplot [color=black, densely dashed]
  table[]{../tiks/datasets/dspartners-15.tsv};

\node[fill=white, below right, align=right, draw=black,anchor=north east, font=\tiny]
at (rel axis cs:0.98,0.97) {$351$};
\end{axis}

\end{tikzpicture}%
%
%
%
\begin{tikzpicture}
\pgfplotsset{
compat=1.11,
legend image code/.code={
\draw[mark repeat=2,mark phase=2]
plot coordinates {
(0cm,0cm)
(0.15cm,0cm)        
(0.3cm,0cm)         
};%
}
}
\begin{axis}[%
xmin=0,
xmax=1400,
xlabel={$t/\unit{s}$},
ymin=278,
ymax=475,
axis background/.style={fill=white},
title style={font=\bfseries},
label style={font=\footnotesize},
legend style={at={(0.02,0.97)},anchor=north west, legend cell align=left, align=left, font=\scriptsize,inner sep=1.5pt},
width = \fwidth, 
 height = \fheight 
]
\addplot [color=black]
  table[]{../tiks/datasets/ds553-1.tsv};

\addplot [color=black, densely dashdotted]
  table[]{../tiks/datasets/ds553-2.tsv};

\addplot [color=black, densely dashed]
  table[]{../tiks/datasets/ds553-3.tsv};

\node[fill=white, below right, align=right, draw=black,anchor=north east, font=\tiny]
at (rel axis cs:0.98,0.97) {$553$};
\end{axis}

\end{tikzpicture}%
%
%
%
\begin{tikzpicture}
\pgfplotsset{
compat=1.11,
legend image code/.code={
\draw[mark repeat=2,mark phase=2]
plot coordinates {
(0cm,0cm)
(0.15cm,0cm)        
(0.3cm,0cm)         
};%
}
}
\begin{axis}[%
xmin=0,
xmax=1400,
xlabel={$t/\unit{s}$},
ymin=278,
ymax=475,
axis background/.style={fill=white},
title style={font=\bfseries},
label style={font=\footnotesize},
legend style={at={(0.02,0.97)},anchor=north west, legend cell align=left, align=left, font=\scriptsize,inner sep=1.5pt},
width = \fwidth, 
 height = \fheight 
]
\addplot [color=black]
  table[]{../tiks/datasets/dspartners-16.tsv};

\addplot [color=black, densely dashdotted]
  table[]{../tiks/datasets/dspartners-17.tsv};

\addplot [color=black, densely dashed]
  table[]{../tiks/datasets/dspartners-18.tsv};

\node[fill=white, below right, align=right, draw=black,anchor=north east, font=\tiny]
at (rel axis cs:0.98,0.97) {$555$};
\end{axis}

\end{tikzpicture}%
%
%
%
\begin{tikzpicture}
\pgfplotsset{
compat=1.11,
legend image code/.code={
\draw[mark repeat=2,mark phase=2]
plot coordinates {
(0cm,0cm)
(0.15cm,0cm)        
(0.3cm,0cm)         
};%
}
}
\begin{axis}[%
xmin=0,
xmax=1400,
xlabel={$t/\unit{s}$},
ymin=278,
ymax=475,
axis background/.style={fill=white},
title style={font=\bfseries},
label style={font=\footnotesize},
legend style={at={(0.02,0.97)},anchor=north west, legend cell align=left, align=left, font=\scriptsize,inner sep=1.5pt},
width = \fwidth, 
 height = \fheight 
]
\addplot [color=black]
  table[]{../tiks/datasets/dspartners-19.tsv};

\addplot [color=black, densely dashdotted]
  table[]{../tiks/datasets/dspartners-20.tsv};

\addplot [color=black, densely dashed]
  table[]{../tiks/datasets/dspartners-21.tsv};

\node[fill=white, below right, align=right, draw=black,anchor=north east, font=\tiny]
at (rel axis cs:0.98,0.97) {$625$};
\end{axis}

\end{tikzpicture}%

%
%
%
\begin{tikzpicture}
\pgfplotsset{
compat=1.11,
legend image code/.code={
\draw[mark repeat=2,mark phase=2]
plot coordinates {
(0cm,0cm)
(0.15cm,0cm)        
(0.3cm,0cm)         
};%
}
}
\begin{axis}[%
xmin=0,
xmax=1400,
xlabel={$t/\unit{s}$},
ymin=278,
ymax=475,
ylabel={$T / \unit{K}$},
axis background/.style={fill=white},
title style={font=\bfseries},
label style={font=\footnotesize},
legend style={at={(0.02,0.97)},anchor=north west, legend cell align=left, align=left, font=\scriptsize,inner sep=1.5pt},
width = \fwidth, 
 height = \fheight 
]
\addplot [color=black]
  table[]{../tiks/datasets/dspartners-22.tsv};

\addplot [color=black, densely dashdotted]
  table[]{../tiks/datasets/dspartners-23.tsv};

\addplot [color=black, densely dashed]
  table[]{../tiks/datasets/dspartners-24.tsv};

\node[fill=white, below right, align=right, draw=black,anchor=north east, font=\tiny]
at (rel axis cs:0.98,0.97) {$638$};
\end{axis}

\end{tikzpicture}%
%
%
%
\begin{tikzpicture}
\pgfplotsset{
compat=1.11,
legend image code/.code={
\draw[mark repeat=2,mark phase=2]
plot coordinates {
(0cm,0cm)
(0.15cm,0cm)        
(0.3cm,0cm)         
};%
}
}
\begin{axis}[%
xmin=0,
xmax=1400,
xlabel={$t/\unit{s}$},
ymin=278,
ymax=475,
axis background/.style={fill=white},
title style={font=\bfseries},
label style={font=\footnotesize},
legend style={at={(0.02,0.97)},anchor=north west, legend cell align=left, align=left, font=\scriptsize,inner sep=1.5pt},
width = \fwidth, 
 height = \fheight 
]
\addplot [color=black]
  table[]{../tiks/datasets/dspartners-25.tsv};

\addplot [color=black, densely dashdotted]
  table[]{../tiks/datasets/dspartners-26.tsv};

\addplot [color=black, densely dashed]
  table[]{../tiks/datasets/dspartners-27.tsv};

\node[fill=white, below right, align=right, draw=black,anchor=north east, font=\tiny]
at (rel axis cs:0.98,0.97) {$692$};
\end{axis}

\end{tikzpicture}%
%
%
%
\begin{tikzpicture}
\pgfplotsset{
compat=1.11,
legend image code/.code={
\draw[mark repeat=2,mark phase=2]
plot coordinates {
(0cm,0cm)
(0.15cm,0cm)        
(0.3cm,0cm)         
};%
}
}
\begin{axis}[%
xmin=0,
xmax=1400,
xlabel={$t/\unit{s}$},
ymin=278,
ymax=475,
axis background/.style={fill=white},
title style={font=\bfseries},
label style={font=\footnotesize},
legend style={at={(0.02,0.97)},anchor=north west, legend cell align=left, align=left, font=\scriptsize,inner sep=1.5pt},
width = \fwidth, 
 height = \fheight 
]
\addplot [color=black]
  table[]{../tiks/datasets/ds745-1.tsv};

\addplot [color=black, densely dashdotted]
  table[]{../tiks/datasets/ds745-2.tsv};

\addplot [color=black, densely dashed]
  table[]{../tiks/datasets/ds745-3.tsv};

\node[fill=white, below right, align=right, draw=black,anchor=north east, font=\tiny]
at (rel axis cs:0.98,0.97) {$745$};
\end{axis}

\end{tikzpicture}%
%
%
%
\begin{tikzpicture}
\pgfplotsset{
compat=1.11,
legend image code/.code={
\draw[mark repeat=2,mark phase=2]
plot coordinates {
(0cm,0cm)
(0.15cm,0cm)        
(0.3cm,0cm)         
};%
}
}
\begin{axis}[%
xmin=0,
xmax=1400,
xlabel={$t/\unit{s}$},
ymin=278,
ymax=475,
axis background/.style={fill=white},
title style={font=\bfseries},
label style={font=\footnotesize},
legend style={at={(0.02,0.97)},anchor=north west, legend cell align=left, align=left, font=\scriptsize,inner sep=1.5pt},
width = \fwidth, 
 height = \fheight 
]
\addplot [color=black]
  table[]{../tiks/datasets/ds795-1.tsv};

\addplot [color=black, densely dashdotted]
  table[]{../tiks/datasets/ds795-2.tsv};

\addplot [color=black, densely dashed]
  table[]{../tiks/datasets/ds795-3.tsv};

\node[fill=white, below right, align=right, draw=black,anchor=north east, font=\tiny]
at (rel axis cs:0.98,0.97) {$795$};
\end{axis}

\end{tikzpicture}%
\end{subfigure}

\caption{Visualization of all discussed data sets, consisting of the excitation signal $T_\text{oven}$ (solid line), core temperature $T_\text{A}$ (dash-dotted line) and surface temperature $T_\text{B}$ (dashed line). Source: Created by author.}
 \label{fig:datasets}
\end{figure*}
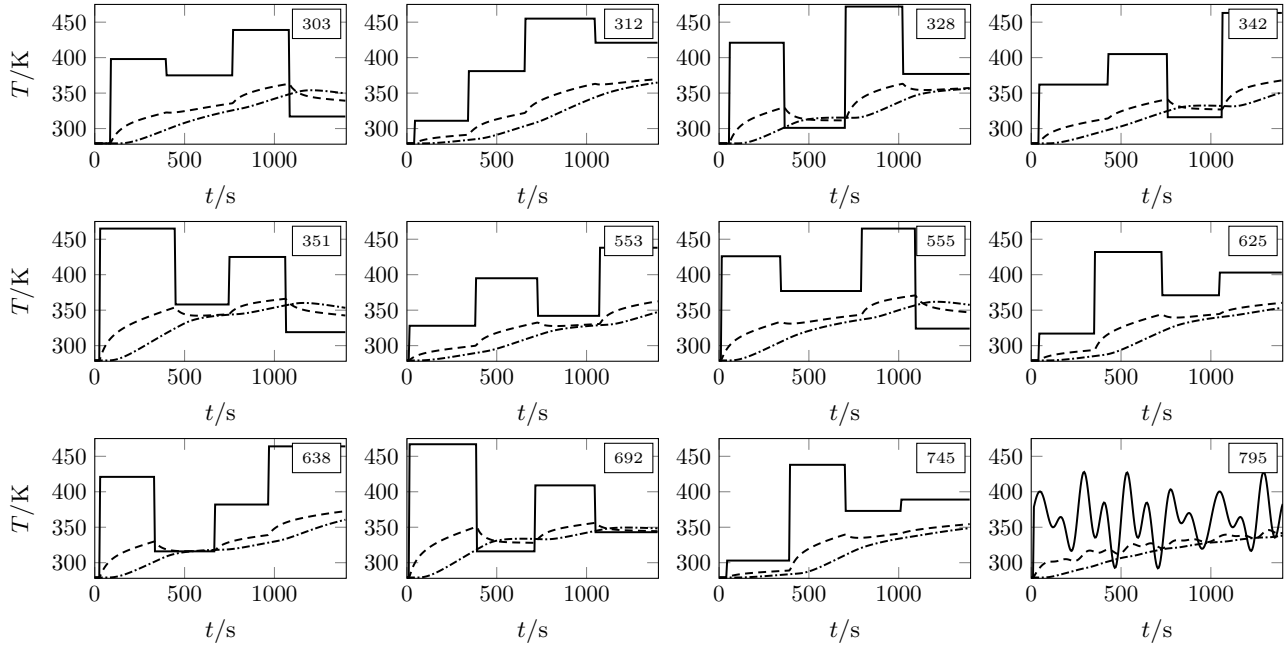

Upon identifying the promising training data set 745, the next step is determining a suitable second training partner data set to enhance ROM training accuracy. 
In this section, we propose a procedure for selecting training partner data sets based on a comparison with the good 1-data-set ROM trained on data set 745.
To evaluate the plausibility of various training partner selection routines, we utilize \autoref{fig:Combo-Comparison}, which incorporates several decision support tools. 
Figure~\ref{fig:combo-a} is a modified version of \autoref{fig:KPIandErrorCorrelations-a}, where the considered data sets are labeled and color-coded. 
Figure~\ref{fig:combo-c} presents $\rmseg$ (bars) and the distribution of $\rmseq$ (boxplots) on AP15 for the \emph{2-data-set ROMs}, whereas \autoref{tab:Eval on APRBSEqual15-Table-Foc-745-combo-comparison} lists all global test errors. 
The subsequent paragraphs introduce hypotheses on how to identify a suitable training partner data set, with consistent naming of the boxes and corresponding color coding in \autoref{fig:Combo-Comparison}.
\textcolor{black}{All discussed data sets are visualized in \autoref{fig:datasets}}.

\begin{table}[t]
\centering
\small
\addtolength{\tabcolsep}{-3.1pt}
 \caption{Performance of 2-data-set ROMs tested on AP15. Source: Created by author.} 
 \begin{tabular}{lrrrrrrrrrrrrrrr} \toprule 
ID & 745 & +553 & +795 & +312& +553+795& +371& +351 & +555 & +638  & +692 & +303 & +328 & +342 & +625  \\ 
Complexity & 2 & 5 & 4 &4& 3& 3 & 4& 3& 5& 4& 3& 3&4&2 \\ 
\midrule 
$\rmseg / \unit{K}$ & 1.05 & \textbf{0.54} & 0.64 & 0.69 & 0.85 & 1.45 & 0.98 & 1.16 & 0.54 & 0.79 & 0.87 & 0.81 & 0.59 & 1.30  \\ 
$\mapeg / \unit{\%}$  & 0.22 & \textbf{0.12} & 0.14 & 0.14 & 0.16 & 0.29 & 0.20 & 0.27 & 0.13 & 0.15 & 0.16 & 0.16 & 0.14 & 0.28  \\ 
$\maxeg / \unit{K}$ & 2.63 & \textbf{1.31} & 1.67 & 1.73 & 2.70 & 3.99 & 2.70 & 2.28 & 1.41 & 2.29 & 2.71 & 2.29 & 1.36 & 3.20  \\ 
$\medeg / \unit{K}$ & 0.07 & \textbf{0.07} & $-$0.19 & 0.02 & 0.03 & 0.01 & 0.13 & $-$0.63 & $-$0.04 & $-$0.07 & $-$0.02 & $-$0.06 & $-$0.12 & 0.01  \\ 
$\iqreg / \unit{K}$  & 1.25 & \textbf{0.62} & 0.64 & 0.73 & 0.80 & 1.31 & 1.05 & 1.43 & 0.51 & 0.73 & 0.75 & 0.80 & 0.72 & 1.44 \\ 
$\overline{\text{R}^2}$ & 1.00 & \textbf{1.00} & 1.00 & 1.00 & 1.00 & 0.99 & 1.00 & 0.99 & 1.00 & 1.00 & 1.00 & 1.00 & 1.00 & 1.00 \\ 
\bottomrule 
 \label{tab:Eval on APRBSEqual15-Table-Foc-745-combo-comparison} \end{tabular}\end{table}

\paragraph{Very similar to 745, best performance on AP15, low ROM745 error:}
One approach involves selecting a dataset as a training partner if it has proven effective for training a 1-data-set ROM. 
The similarity between an excitation signal $j$ and APRBS 745 is quantified by $\operatorname{rms}(\disphat{T}_\text{oven,745} - \disphat{T}_{\text{oven,}j}) $, represented as the ordinate in \autoref{fig:combo-b}. 
Another criterion for training partner selection is the test error of the 1-data-set ROM745 on each individual APRBS dataset, plotted as the abscissa in \autoref{fig:combo-b}. 
Notably, the 1-data-set ROM745 exhibits low test errors on the 625 test set. 
This aligns with the similarity between data set 625 (color-coded in black) and data set 745 regarding oven temperatures, evident from the low $y$-position in \autoref{fig:combo-b}. 
Additionally, the 1-data-set ROM625 performs exceptionally well on AP15, as shown in \autoref{fig:combo-a}. 
However, selecting data sets 745 and 625 as training partners to form ROM745+625 proves disadvantageous, leading to an increase in test errors instead of a decrease, as evident in the black section of \autoref{fig:combo-c}. 
Data set 625 does not contribute new information to the training process; instead, it may induce overfitting of the ROM to the specific, similar operational conditions of data sets 745 and 625.

\paragraph{Somewhat similar to 745, best performance on AP15, medium ROM745 error:}
Despite the example above, combining training data sets that independently train accurate 1-data-set ROMs remains a plausible approach. 
However, a certain level of dissimilarity and a medium test error of the base ROM745 on those datasets should be ensured, as indicated by the red color-coded cases in \autoref{fig:Combo-Comparison}. 
The ROM745+553 achieves a \qty{49}{\percent} reduction in test error to $\rmseg = \SI{0.54}{K}$. It exhibits a significantly low error spread in replicating the full-order model --- the best result in this study on AP15. 
Additionally, combining data set 745 with a moderately similar but best-performing multi-sine 1-data-set ROM795 training data set yields decent results of $\rmseg = \SI{0.64}{K}$ on AP15. 
However, combining all three signals, as demonstrated in the case ROM745+795+553, does not enhance model quality.

\paragraph{Dissimilar to 745, medium or low performance on AP15, medium ROM745 error:}
Another intuitive approach is to select a training partner that differs from the base data set 745 in terms of oven temperature. 
However, improving ROM quality is not guaranteed with this method. 
Highly dissimilar data sets with moderate performance as 1-data-set ROMs (color-coded in blue) or dissimilar data sets that train poor 1-data-set ROMs (color-coded in light gray) do not significantly enhance ROM quality.

\paragraph{Somewhat similar to 745, medium performance on AP15, high ROM745 error:}
An effective strategy is to add data sets as training partners where $\rmseq$ of ROM745 is high (color-coded in green). 
The decreasing trend of test error medians for the light gray and green training signals (see the trend arrows in \autoref{fig:Combo-Comparison}) suggests that training improvement correlates with increasing test error of 1-data-set ROM745 (abscissa in \autoref{fig:combo-b}).

\bigskip

\textcolor{black}{The best training data combination can be found with the help of
Figure~\ref{fig:combo-b}, which can be interpreted as a similarity map of data sets.}
Placing the second training data set close to a cluster of points seems to enrich ROM training. 
However, as discussed in the above hypotheses, this does not automatically guarantee effective training data combinations.
Based on this study, a training partner data set should at least yield good 1-data-set ROMs and exhibit a certain level of dissimilarity to the base training data set regarding oven temperatures. \textcolor{black}{As indicated by the arrows in \autoref{fig:Combo-Comparison},} the base-signal ROM should have elevated test errors on the potential partner data set. 
Suitable training partner candidates are identified along the main diagonal of \autoref{fig:combo-b} (red and green cases).  \textcolor{black}{This finding features potential follow-up research to classify potential training partners more granularly}. 

%


%

\begin{figure*}[t] 
\centering
\captionsetup[subfigure]{aboveskip=-1pt,belowskip=5pt}
\begin{subfigure}[]{\textwidth}
\centering
\setlength\fheight{0.15\textheight}
\setlength\fwidth{0.28\textwidth}
%
%
\definecolor{mycolor1}{rgb}{0.00000,0.61176,0.85490}%
\definecolor{mycolor2}{rgb}{0.65098,0.00000,0.51765}%
\definecolor{mycolor3}{rgb}{0.96078,0.63922,0.00000}%
\definecolor{mycolor4}{rgb}{1.00000,0.00000,1.00000}%
\definecolor{mycolor5}{rgb}{1.00000,0.84314,0.00000}%
\begin{tikzpicture}
\pgfplotsset{
compat=1.11,
legend image code/.code={
\draw[mark repeat=2,mark phase=2]
plot coordinates {
(0cm,0cm)
(0.15cm,0cm)        
(0.3cm,0cm)         
};%
}
}
\begin{axis}[%
xmin=0,
xmax=1400,
xlabel={$t/\unit{s}$},
ymin=278,
ymax=465,
ylabel={$T_\text{oven} / \unit{K}$},
axis background/.style={fill=white},
title style={font=\bfseries},
label style={font=\footnotesize},
legend style={at={(0.02,0.97)},anchor=north west, legend cell align=left, align=left, font=\scriptsize,inner sep=1.5pt},
width = \fwidth, 
 height = \fheight 
]
%
%
\addplot [color=black]
  table[]{../tiks/ROM-signal/TimeEval-840sines-3.tsv};
\addlegendentry{$840$}

%

%
%
%
%
%

\end{axis}

\begin{axis}[%
xmin=0,
xmax=1,
ymin=0,
ymax=1,
axis line style={only marks},
ticks=none,
axis x line*=bottom,
axis y line*=left,
legend style={legend cell align=left, align=left, draw=white!15!black},
width = \fwidth, 
 height = \fheight 
]
\end{axis}
\end{tikzpicture}%
%
%
\definecolor{mycolor1}{rgb}{0.00000,0.61176,0.85490}%
\definecolor{mycolor2}{rgb}{0.75294,0.75294,0.75294}%
\definecolor{mycolor3}{rgb}{0.00000,0.49804,0.00000}%
\definecolor{mycolor4}{rgb}{0.96078,0.63922,0.00000}%
\definecolor{mycolor5}{rgb}{0.65098,0.00000,0.51765}%
\begin{tikzpicture}
\pgfplotsset{
compat=1.11,
legend image code/.code={
\draw[mark repeat=2,mark phase=2]
plot coordinates {
(0cm,0cm)
(0.15cm,0cm)        
(0.3cm,0cm)         
};%
}
}
\begin{axis}[%
xmin=0,
xmax=1400,
xlabel={$t/\unit{s}$},
ymin=278,
ymax=465,
label style={font=\footnotesize},
axis background/.style={fill=white},
title style={font=\bfseries},
legend style={at={(0.02,0.97)},anchor=north west, legend cell align=left, align=left,  font=\scriptsize, inner sep=1.5pt},
width = \fwidth, 
 height = \fheight 
]
%

\addplot [color=black]
  table[]{../tiks/ROM-signal/TimeEval-303sines-3.tsv};
\addlegendentry{$303$}

%
%
%
%
%

\end{axis}

\begin{axis}[%
xmin=0,
xmax=1,
ymin=0,
ymax=1,
axis line style={only marks},
ticks=none,
axis x line*=bottom,
axis y line*=left,
legend style={legend cell align=left, align=left, draw=white!15!black},
width = \fwidth, 
 height = \fheight 
]
\end{axis}
\end{tikzpicture}%
%
%
\definecolor{mycolor1}{rgb}{0.00000,0.61176,0.85490}%
\definecolor{mycolor2}{rgb}{0.00000,1.00000,1.00000}%
\definecolor{mycolor3}{rgb}{0.96078,0.63922,0.00000}%
\definecolor{mycolor4}{rgb}{0.65098,0.00000,0.51765}%
\begin{tikzpicture}
\pgfplotsset{
compat=1.11,
legend image code/.code={
\draw[mark repeat=2,mark phase=2]
plot coordinates {
(0cm,0cm)
(0.15cm,0cm)        
(0.3cm,0cm)         
};%
}
}
\begin{axis}[%
xmin=0,
xmax=1400,
xlabel={$t/\unit{s}$},
ymin=278,
ymax=465,
axis background/.style={fill=white},
title style={font=\bfseries},
label style={font=\footnotesize},
legend style={at={(0.02,0.97)},anchor=north west, legend cell align=left, align=left,  font=\scriptsize,inner sep=1.5pt},
width = \fwidth, 
 height = \fheight 
]
%
%
%
%
\addplot [color=black]
  table[]{../tiks/ROM-signal/TimeEval-703sines-5.tsv};
\addlegendentry{$703$}
%
%
%
%

\end{axis}

\begin{axis}[%
xmin=0,
xmax=1,
ymin=0,
ymax=1,
axis line style={only marks},
ticks=none,
axis x line*=bottom,
axis y line*=left,
legend style={legend cell align=left, align=left, draw=white!15!black},
width = \fwidth, 
 height = \fheight 
]
\end{axis}
\end{tikzpicture}%
%
%
\definecolor{mycolor1}{rgb}{0.00000,0.61176,0.85490}%
\definecolor{mycolor2}{rgb}{0.65098,0.00000,0.51765}%
\definecolor{mycolor3}{rgb}{0.96078,0.63922,0.00000}%
\begin{tikzpicture}
\pgfplotsset{
compat=1.11,
legend image code/.code={
\draw[mark repeat=2,mark phase=2]
plot coordinates {
(0cm,0cm)
(0.15cm,0cm)        
(0.3cm,0cm)         
};%
}
}
\begin{axis}[%
xmin=0,
xmax=1400,
xlabel style={font=\color{white!15!black}},
xlabel={$t/\unit{s}$},
ymin=278,
ymax=465,
ylabel style={font=\color{white!15!black}},
label style={font=\footnotesize},
axis background/.style={fill=white},
legend style={at={(0.02,0.97)},anchor=north west, legend cell align=left, align=left, font=\scriptsize,inner sep=1.5pt},
width = \fwidth, 
 height = \fheight 
]
\addplot [color=black]
  table[]{../tiks/ROM-signal/TimeEval-1050sines-1.tsv};
\addlegendentry{$1050$}

%
%
%
%
%
%
%

\end{axis}

\begin{axis}[%
xmin=0,
xmax=1,
ymin=0,
ymax=1,
axis line style={only marks},
ticks=none,
axis x line*=bottom,
axis y line*=left,
legend style={legend cell align=left, align=left, draw=white!15!black},
width = \fwidth, 
 height = \fheight 
]
\end{axis}
\end{tikzpicture}%
\caption{Oven temperatures of test data sets.}\label{fig:timeevals-a}
\vspace{0.6em}
\end{subfigure}
\begin{subfigure}[]{\textwidth}
\centering
\setlength\fheight{0.20\textheight}
\setlength\fwidth{0.28\textwidth} 
%
%
\definecolor{mycolor1}{rgb}{0.00000,0.61176,0.85490}%
\definecolor{mycolor2}{rgb}{0.65098,0.00000,0.51765}%
\definecolor{mycolor3}{rgb}{0.96078,0.63922,0.00000}%
\begin{tikzpicture}
\pgfplotsset{
compat=1.11,
legend image code/.code={
\draw[mark repeat=2,mark phase=2]
plot coordinates {
(0cm,0cm)
(0.15cm,0cm)        
(0.3cm,0cm)         
};%
}
}
\begin{axis}[%
xmin=0,
xmax=1400,
xlabel={$t/\unit{s}$},
ymin=278,
ymax=369,
ylabel={$T /\unit{K}$},
axis background/.style={fill=white},
label style={font=\footnotesize},
legend style={at={(0.5,1.02)},anchor=south, legend cell align=left, align=left},
legend style={nodes={scale=0.752, transform shape}}, 
width = \fwidth, 
 height = \fheight 
]
\addplot [color=mygrey, forget plot]
  table[]{../tiks/ROM-signal/TimeEval-840bestcombo-1.tsv};
\addlegendentry{$\text{840-sr3nl473h300t1400 - T}_\text{a}\text{ / K}$}

\addplot [color=tud2a, opacity=0.65,forget plot]
  table[]{../tiks/ROM-signal/TimeEval-840bestcombo-2.tsv};
\addlegendentry{$\text{840-sr3nl473h300t1400 - T}_\text{b}\text{ / K}$}
\legend{}
\addplot [color=black, dashdotted]
  table[]{../tiks/ROM-signal/TimeEval-840bestcombo-3.tsv};
\addlegendentry{745+795 $T_\text{A}$: $\rmseq$ = 0.3 K}

\addplot [color=black, dashed]
  table[]{../tiks/ROM-signal/TimeEval-840bestcombo-4.tsv};
\addlegendentry{745+795 $T_\text{B}$: $\rmseq$ = 0.2 K} 

\end{axis}

\begin{axis}[%
xmin=0,
xmax=1,
ymin=0,
ymax=1,
axis line style={only marks},
ticks=none,
axis x line*=bottom,
axis y line*=left,
legend style={legend cell align=left, align=left, draw=white!15!black},
width = \fwidth, 
 height = \fheight 
]
\end{axis}
\end{tikzpicture}%
%
%
\definecolor{mycolor1}{rgb}{0.00000,0.61176,0.85490}%
\definecolor{mycolor2}{rgb}{0.65098,0.00000,0.51765}%
\definecolor{mycolor3}{rgb}{0.96078,0.63922,0.00000}%
\begin{tikzpicture}
\pgfplotsset{
compat=1.11,
legend image code/.code={
\draw[mark repeat=2,mark phase=2]
plot coordinates {
(0cm,0cm)
(0.15cm,0cm)        
(0.3cm,0cm)         
};%
}
}
\begin{axis}[%
xmin=0,
xmax=1400,
xlabel={$t/\unit{s}$},
ymin=278,
ymax=369,
axis background/.style={fill=white},
title style={font=\bfseries},
label style={font=\footnotesize},
legend style={at={(0.5,1.02)},anchor=south, legend cell align=left, align=left},
legend style={nodes={scale=0.752, transform shape}}, 
width = \fwidth, 
 height = \fheight 
]
\addplot [color=mygrey, , forget plot]
  table[]{../tiks/ROM-signal/TimeEval-303bestcombo-1.tsv};
\addlegendentry{$\text{T}_\text{a}\text{ / K}$}

\addplot [color=tud2a, opacity=0.65, forget plot]
  table[]{../tiks/ROM-signal/TimeEval-303bestcombo-2.tsv};
\addlegendentry{$\text{T}_\text{b}\text{ / K}$}
\legend{}
\addplot [color=black, dashdotted]
  table[]{../tiks/ROM-signal/TimeEval-303bestcombo-3.tsv};
\addlegendentry{745+795 $T_\text{A}$: $\rmseq$ = 0.2 K} 

\addplot [color=black, dashed]
  table[]{../tiks/ROM-signal/TimeEval-303bestcombo-4.tsv};
\addlegendentry{745+795 $T_\text{B}$: $\rmseq$ = 0.7 K} 

\end{axis}

\begin{axis}[%
xmin=0,
xmax=1,
ymin=0,
ymax=1,
axis line style={only marks},
ticks=none,
axis x line*=bottom,
axis y line*=left,
legend style={legend cell align=left, align=left, draw=white!15!black},
width = \fwidth, 
 height = \fheight 
]
\end{axis}
\end{tikzpicture}%
%
%
\definecolor{mycolor1}{rgb}{0.00000,0.61176,0.85490}%
\definecolor{mycolor2}{rgb}{0.65098,0.00000,0.51765}%
\definecolor{mycolor3}{rgb}{0.96078,0.63922,0.00000}%
\begin{tikzpicture}
\pgfplotsset{
compat=1.11,
legend image code/.code={
\draw[mark repeat=2,mark phase=2]
plot coordinates {
(0cm,0cm)
(0.15cm,0cm)        
(0.3cm,0cm)         
};%
}
}
\begin{axis}[%
xmin=0,
xmax=1400,
xlabel={$t/\unit{s}$},
ymin=278,
ymax=369,
axis background/.style={fill=white},
title style={font=\bfseries},
label style={font=\footnotesize},
legend style={at={(0.5,1.02)},anchor=south, legend cell align=left, align=left},
legend style={nodes={scale=0.752, transform shape}}, 
width = \fwidth, 
 height = \fheight 
]
\addplot [color=mygrey,, forget plot ]
  table[]{../tiks/ROM-signal/TimeEval-703bestcombo-1.tsv};
\addlegendentry{$\text{T}_\text{a}\text{ / K}$}

\addplot [color=tud2a, opacity=0.65, forget plot]
  table[]{../tiks/ROM-signal/TimeEval-703bestcombo-2.tsv};
\addlegendentry{$\text{T}_\text{b}\text{ / K}$}
\legend{}
\addplot [color=black, dashdotted]
  table[]{../tiks/ROM-signal/TimeEval-703bestcombo-3.tsv};
\addlegendentry{745+795 $T_\text{A}$: $\rmseq$ = 0.2 K}   

\addplot [color=black, dashed]
  table[]{../tiks/ROM-signal/TimeEval-703bestcombo-4.tsv};
\addlegendentry{745+795 $T_\text{B}$: $\rmseq$ = 0.2 K} 

\end{axis}

\begin{axis}[%
xmin=0,
xmax=1,
ymin=0,
ymax=1,
axis line style={only marks},
ticks=none,
axis x line*=bottom,
axis y line*=left,
legend style={legend cell align=left, align=left, draw=white!15!black},
width = \fwidth, 
 height = \fheight 
]
\end{axis}
\end{tikzpicture}%
%
%
\definecolor{mycolor1}{rgb}{0.00000,0.61176,0.85490}%
\definecolor{mycolor2}{rgb}{0.65098,0.00000,0.51765}%
\definecolor{mycolor3}{rgb}{0.96078,0.63922,0.00000}%
\begin{tikzpicture}
\pgfplotsset{
compat=1.11,
legend image code/.code={
\draw[mark repeat=2,mark phase=2]
plot coordinates {
(0cm,0cm)
(0.15cm,0cm)        
(0.3cm,0cm)         
};%
}
}
\begin{axis}[%
xmin=0,
xmax=1400,
xlabel={$t/\unit{s}$},
ymin=278,
ymax=369,
axis background/.style={fill=white},
title style={font=\bfseries},
label style={font=\footnotesize},
legend style={at={(0.5,1.02)},anchor=south, legend cell align=left, align=left},
legend style={nodes={scale=0.752, transform shape}}, 
width = \fwidth, 
 height = \fheight 
]
\addplot [color=mygrey, forget plot ]
  table[]{../tiks/ROM-signal/TimeEval-1050bestcombo-1.tsv};
\addlegendentry{$\text{T}_\text{a}\text{ / K}$}

\addplot [color=tud2a, opacity=0.65, forget plot]
  table[]{../tiks/ROM-signal/TimeEval-1050bestcombo-2.tsv};
\addlegendentry{$\text{T}_\text{b}\text{ / K}$}
\legend{}
\addplot [color=black, dashdotted]
  table[]{../tiks/ROM-signal/TimeEval-1050bestcombo-3.tsv};
\addlegendentry{745+795 $T_\text{A}$: $\rmseq$ = 0.3 K} 

\addplot [color=black, dashed]
  table[]{../tiks/ROM-signal/TimeEval-1050bestcombo-4.tsv};
\addlegendentry{745+795 $T_\text{B}$: $\rmseq$ = 0.4 K} 

\end{axis}

\begin{axis}[%
xmin=0,
xmax=1,
ymin=0,
ymax=1,
axis line style={only marks},
ticks=none,
axis x line*=bottom,
axis y line*=left,
legend style={legend cell align=left, align=left, draw=white!15!black},
width = \fwidth, 
 height = \fheight 
]
\end{axis}
\end{tikzpicture}\caption{Core temperature evolutions of the test data sets.}\label{fig:timeevals-d}
\end{subfigure}
\caption{Time evaluation of the representative ROM745+795 on four test data sets. Grey and blue solid lines represent the full-order model solutions $T_\text{A}$ and $T_\text{B}$, while dashed lines indicate the ROM predictions. Source: Created by author.}
 \label{fig:timeevals}
\end{figure*}
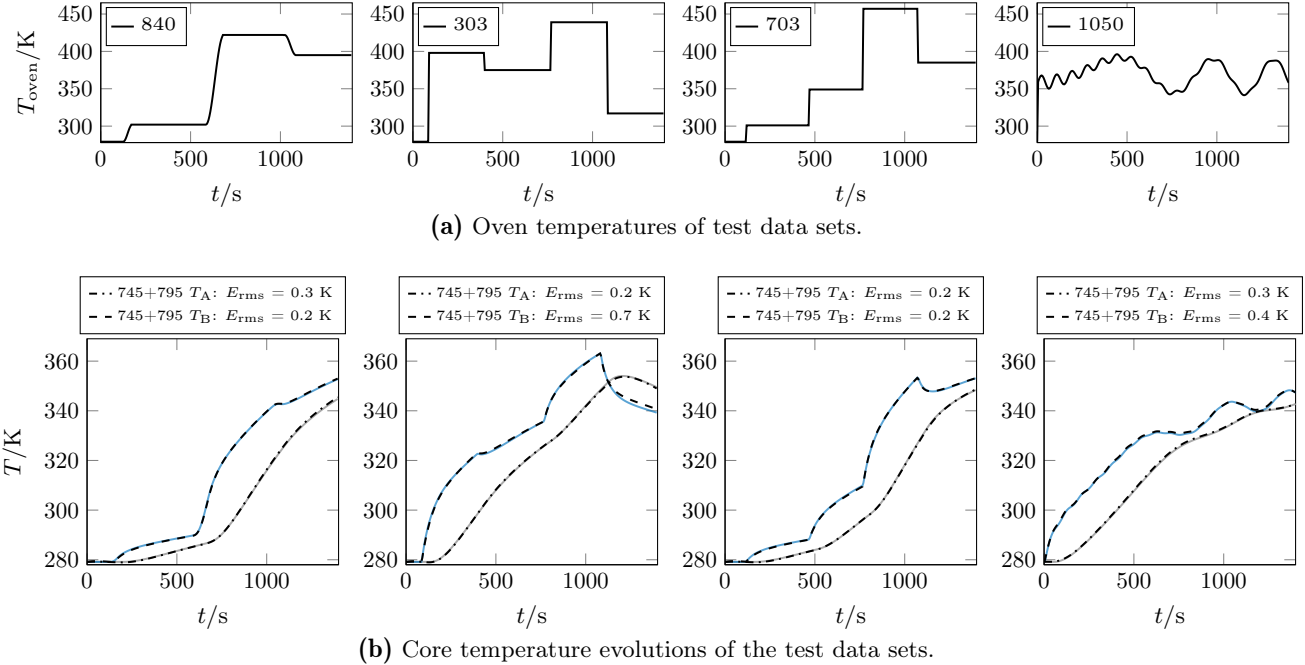

The 2-data-set ROM745+553 performs best on AP15. Among various test groups that also feature sinusoidal test signals, the 2-data-set ROM745+795 exhibits slightly superior performance. 
It achieves a low global test error of $\rmseg = \SI{0.64}{K}$ on AP15, $\rmseg = \SI{0.38}{K}$ on 15 selected multi-sine test data, and $\rmseg = \SI{0.67}{K}$ on 15 APRBS signals with sinusoidal transitions, referred to as sinAPRBS~\autocite{rom_Nelles2020}. 
For a comprehensive evaluation of the errors of all 2-data-set ROMs on these and additional test groups, the reader is directed to \autocite{diss_phd23}.
The time evaluation of the 2-data-set ROM745+795 is illustrated in \autoref{fig:timeevals-d}, revealing a characteristic $\rmseq < \SI{0.5}{K}$ for the presented test data. 

\textcolor{black}{The full finite element solutions of the presented case (a quarter section of the food model) take about five hours on 20 cores of a cluster of two Intel Xeon E5-2687W v4 (3.2 GHz) processors. In comparison, the reduced order model requires approximately $\SI{0.10}{s}$ to predict one hour of real time, which translates to a speed-up of $\operatorname{Sp} \approx \SI{3.6E4}{}$ with no noticeable computational load on a single processor of the same computer.}

\subsection{Discussion of the proposed design of experiments}
The proposed efficient design of experiments addresses the challenge of selecting a limited number of appropriate training data sets for accurate ROMs. 
This is achieved by correlating training data features with error measures of 1-data-set ROMs, allowing for the identification or generation of optimal training data.

When choosing a suitable excitation signal type, it is crucial to differentiate between ROM training, testing, and operation. 
In this study, ROM training proves to be most successful with data sets excited by APRBS. 
However, it is worth noting that a real convection oven may not replicate the instantaneous jumps in oven temperature characteristic of APRBS signals. 
This observation challenges conventional recommendations in the literature, which often advocate using signal types closely resembling operational signals for training purposes \autocite{sig_Gringard2016}.

Testing must be conducted on objective test data sets to ensure the robustness and generalization of the reduced-order models. 
Randomly generating or selecting training data, as seen in methods like $k$-fold cross-validation, might not uniformly represent the output space at all positions, as discussed in detail in \autocite{diss_phd23}.
One can employ methods such as selecting data sets based on a $\chi^2$ test to address this issue and achieve a more balanced representation of the output space. 
This ensures, for example, a uniform distribution of the medians of the output variable. 
Fair test groups are essential for uncovering correlations between data set features and error measures.
Additionally, testing can be performed using different excitation signal classes. 
While APRBS with sinusoidal transitions (sinAPRBS) most accurately emulate convection oven operating temperatures, testing on various signal classes provides a comprehensive evaluation. 
It is worth noting that testing on APRBS is considered conservative, as ROMs trained on APRBS data tend to exhibit even lower test errors on sinAPRBS test data in this study.

It is important to note that low correlation values in the Pearson correlation matrix only imply the absence of a \emph{linear} correlation. 
Nonlinear dependencies between data set features and error measures might still exist. 
This limitation is somewhat alleviated in this work by selecting error measures that exhibit linear or quadratic behavior in either a local or global sense. 
Future research could explore the derivation of other nonlinear data set features.
Even if no correlation is identified between any feature and error measure, the pre-search efforts are valuable. 
The parameter sweep conducted aids in selecting the best training data set available. 
Additionally, regardless of whether a correlation is established, the similarity chart for the best 1-data-set ROM facilitates the selection of additional data sets to enhance ROM accuracy.
For instance, a suitable partner data set should demonstrate dissimilarity to the base data set, and elevated mutual 1-data-set ROM test error levels should be considered when selecting potential training partners. 
Moreover, all viable training partners should be capable of generating good 1-data-set ROMs.
\textcolor{black}{All in all, correlations are strongly problem-related. That is plausible, as different physical models would require different characteristics in the training data to ensure rich information content in the data set.
A recent publication~\autocite{diss_phd23} demonstrates how TwinLab's design of experiments method flexibly adapts to the physics of other models. }

\section{Conclusion}\label{sec:conclusion}
The research presented in this work underscores the potency of data-driven reduced-order modeling in crafting precise and efficient digital twins from simulation data. 
The methodology is agnostic to modeling software, offering versatility across various platforms. 
Remarkably, data-driven ROMs of high accuracy and speed-up can be developed with just two training data sets. 
The proposed efficient design of experiments facilitates the selection of suitable training data sets.
The ROMs introduced in this work exhibit substantial speed-ups compared to real time, achieving approximately $\operatorname{Sp} \approx \SI{3.6E4}{}$ with characteristic solution times of one-tenth of a second without imposing a noticeable computational cost on one core of a processor. 
Beyond the notable speed-up, the data-driven ROMs demonstrate excellent test accuracy across large and representative test data groups.
Specifically, root-mean-square errors for the best point data ROM range from $\qtyrange{0.30}{0.74}{K}$, varying depending on the test group. 
\textcolor{black}{The fusion of high accuracy with computing efficiency enables the digital twin to run directly on the control logic of the physical process, eliminating the need for edge or cloud computing.}

\subsection{Contributions} 

The TwinLab framework is a valuable tool for automating the generation of data-driven ROMs from simulation data. 
It streamlines excitation signal generation, test group selection based on $\chi^2$ tests, determination of data set features, and calculation of global test errors. 
The framework also incorporates code for batch simulations of full-order models in COMSOL Multiphysics or ANSYS Fluent, effectively reducing data handling and user-interaction efforts.
\textcolor{black}{The proposed design of experiments may help less experienced users select appropriate training data for reduced-order modeling.}

\subsection{Outlook}
This study illustrates the practical application of physics-based, data-driven digital twins in thermal food processing. 
In a broader context, the digital twin framework contributes to a paradigm shift in the perspective on digital twins by emphasizing process autonomy. 
In the first future steps, the framework's versatility may extend to other domains where the application aligns closely with physics principles. 
For instance, the optimal operation of fuel cells, which involves modeling porous media, fluid flow, heat exchange, and additional reaction kinetics, is strongly correlated to physics. Anticipated developments in mobile applications, particularly within the transportation sector, could leverage digital twins to intelligently control and achieve optimal operating conditions for fuel cells.
Moreover, future work with the proposed framework may extend beyond applications within the presented physics and can extend its contribution to various fields with different physical characteristics. 
For example, in additive manufacturing~\autocite{valizadeh2023}, where the physical process parameters are decisive for end quality, the framework can enhance process autonomy by providing physics-based, data-driven digital twins.

\section*{Acknowledgments}

The authors acknowledge the support of the Graduate School CE within the Centre for Computational Engineering at the Technical University of Darmstadt and thank Minh Khang Pham for generating some of the employed data sets. 



\appendix





\renewcommand*{\bibfont}{\footnotesize}
\printbibliography

@article{dt_moyaDigitalTwinsThat2022,
	author = {Moya, Beatriz and Bad{\'\i}as, Alberto and Alfaro, Ic{\'\i}ar and Chinesta, Francisco and Cueto, El{\'\i}as},
	doi = {10.1002/nme.6535},
	issn = {1097-0207},
	journal = {International Journal for Numerical Methods in Engineering},
	language = {en},
	number = {13},
	pages = {3034--3044},
	title = {Digital twins that learn and correct themselves},
	url = {https://onlinelibrary.wiley.com/doi/abs/10.1002/nme.6535},
	urldate = {2023-11-17},
	volume = {123},
	year = {2022}}

@article{rom_guenotAdaptiveSamplingStrategies2013,
	author = {Gu{\'e}not, Marc and Lepot, Ingrid and Sainvitu, Caroline and Goblet, Jordan and Filomeno Coelho, Rajan},
	doi = {10.1108/02644401311329352},
	editor = {Vasile, Massimiliano and Minisci, Edmondo and Quagliarella, Domenico},
	issn = {0264-4401},
	journal = {Engineering Computations},
	month = jan,
	note = {Publisher: Emerald Group Publishing Limited},
	number = {4},
	pages = {521--547},
	title = {Adaptive sampling strategies for non‐intrusive {POD}‐based surrogates},
	url = {https://doi.org/10.1108/02644401311329352},
	urldate = {2023-11-17},
	volume = {30},
	year = {2013}}

@phdthesis{diss_phd23,
	author = {Maximilian Kannapinn},
	doi = {10.26083/tuprints-00024386},
	school = {Technical University of Darmstadt},
	title = {Digital Twins for Autonomous Thermal Food Processing},
	year = {2023}}

@article{rom_Rivas2013,
	author = {Rivas, Diana and Vilas, Carlos and Alonso, Antonio A. and Varas, Fernando},
	doi = {10.1111/jfpe.12010},
	journal = {Journal of Food Process Engineering},
	number = {4},
	pages = {480-491},
	title = {Derivation of Postharvest Fruit Behavior Reduced Order Models for Online Monitoring and Control of Quality Parameters During Refrigeration},
	url = {https://onlinelibrary.wiley.com/doi/abs/10.1111/jfpe.12010},
	volume = {36},
	year = {2013}}

@article{rom_pehersdorfer2016,
	author = {Benjamin Peherstorfer and Karen Willcox},
	doi = {10.1016/j.cma.2016.03.025},
	issn = {0045-7825},
	journal = {Computer Methods in Applied Mechanics and Engineering},
	pages = {196-215},
	title = {Data-driven operator inference for nonintrusive projection-based model reduction},
	url = {https://www.sciencedirect.com/science/article/pii/S0045782516301104},
	volume = {306},
	year = {2016}}

@article{dt_shrivastava2023twinpackage,
	author = {Chandrima Shrivastava and Seraina Schudel and Kanaha Shoji and Daniel Onwude and F{\'a}tima Pereira {da Silva} and Deniz Turan and Maxence Paillart and Thijs Defraeye},
	doi = {10.1016/j.postharvbio.2023.112283},
	issn = {0925-5214},
	journal = {Postharvest Biology and Technology},
	pages = {112283},
	title = {Digital twins for selecting the optimal ventilated strawberry packaging based on the unique hygrothermal conditions of a shipment from farm to retailer},
	url = {https://www.sciencedirect.com/science/article/pii/S0925521423000443},
	volume = {199},
	year = {2023}}

@article{dt_Defraeye2021coming,
	author = {Thijs Defraeye and Chandrima Shrivastava and Tarl Berry and Pieter Verboven and Daniel Onwude and Seraina Schudel and Andreas B{\"u}hlmann and Paul Cronje and Ren{\'e} M. Rossi},
	doi = {10.1016/j.tifs.2021.01.025},
	issn = {0924-2244},
	journal = {Trends in Food Science \& Technology},
	pages = {245-258},
	title = {Digital twins are coming: {Will} we need them in supply chains of fresh horticultural produce?},
	url = {https://www.sciencedirect.com/science/article/pii/S092422442100025X},
	volume = {109},
	year = {2021}}

@article{rom_kim2022dynromwind,
	article-number = {6329},
	author = {Kim, Changhyun and Dinh, Minh-Chau and Sung, Hae-Jin and Kim, Kyong-Hwan and Choi, Jeong-Ho and Graber, Lukas and Yu, In-Keun and Park, Minwon},
	doi = {10.3390/en15176329},
	issn = {1996-1073},
	journal = {Energies},
	number = {17},
	pages = {6329},
	title = {Design, Implementation, and Evaluation of an Output Prediction Model of the 10 {MW} Floating Offshore Wind Turbine for a Digital Twin},
	url = {https://www.mdpi.com/1996-1073/15/17/6329},
	volume = {15},
	year = {2022}}

@article{rom_Boscaglia2021,
	author = {Boscaglia, Luca and Boglietti, Aldo and Nategh, Shafigh and Bonsanto, Fabio and Scema, Claudio},
	doi = {10.1109/TIA.2021.3077553},
	journal = {IEEE Transactions on Industry Applications},
	number = {4},
	pages = {4118-4129},
	title = {Numerically Based Reduced-Order Thermal Modeling of Traction Motors},
	volume = {57},
	year = {2021}}

@article{rom_broyart2003rnn,
	author = {B. Broyart and G. Trystram},
	doi = {10.1205/096030803322756402},
	issn = {0960-3085},
	journal = {Food and Bioproducts Processing},
	number = {4},
	pages = {316-326},
	title = {Modelling of Heat and Mass Transfer Phenomena and Quality Changes During Continuous Biscuit Baking Using Both Deductive and Inductive (Neural Network) Modelling Principles},
	url = {https://www.sciencedirect.com/science/article/pii/S0960308503703939},
	volume = {81},
	year = {2003}}

@article{rom_Isleroglu2020predictioncookieANN,
	author = {Isleroglu, Hilal and Beyhan, Selami},
	date = {2020/07/01},
	doi = {10.1007/s00231-020-02837-6},
	id = {Isleroglu2020},
	isbn = {1432-1181},
	journal = {Heat and Mass Transfer},
	number = {7},
	pages = {2045--2055},
	title = {Prediction of baking quality using machine learning based intelligent models},
	url = {https://doi.org/10.1007/s00231-020-02837-6},
	volume = {56},
	year = {2020}}

@article{rom_Lu21node-vs-lstm-pharma,
	author = {James Lu and Kaiwen Deng and Xinyuan Zhang and Gengbo Liu and Yuanfang Guan},
	doi = {10.1016/j.isci.2021.102804},
	issn = {2589-0042},
	journal = {iScience},
	number = {7},
	pages = {102804},
	title = {{Neural-ODE} for pharmacokinetics modeling and its advantage to alternative machine learning models in predicting new dosing regimens},
	url = {https://www.sciencedirect.com/science/article/pii/S2589004221007720},
	volume = {24},
	year = {2021}}

@article{rom_pepe22node-vs-lstm-battery,
	author = {Simona Pepe and Jiapeng Liu and Emanuele Quattrocchi and Francesco Ciucci},
	doi = {10.1016/j.est.2022.104209},
	issn = {2352-152X},
	journal = {Journal of Energy Storage},
	pages = {104209},
	title = {Neural ordinary differential equations and recurrent neural networks for predicting the state of health of batteries},
	url = {https://www.sciencedirect.com/science/article/pii/S2352152X22002407},
	volume = {50},
	year = {2022}}

@misc{diss_twinlab,
	author = {Maximilian Kannapinn},
	howpublished = {https://github.com/maxkann/twinlab},
	title = {{TwinLab} -- {A} {MATLAB} code framework for digital twin generation from {COMSOL} and {ANSYS} simulation data},
	year = {2023}}

@book{rom_brunton22databook,
	author = {Steven L. Brunton and J. Nathan Kutz},
	edition = {2nd},
	publisher = {Cambridge University Press},
	title = {Data-Driven Science and Engineering: {Machine} Learning, Dynamical Systems, and Control},
	year = {2022}}

@inproceedings{rom_he2016resnet,
	author = {He, Kaiming and Zhang, Xiangyu and Ren, Shaoqing and Sun, Jian},
	booktitle = {2016 IEEE Conference on Computer Vision and Pattern Recognition (CVPR)},
	doi = {10.1109/CVPR.2016.90},
	pages = {770-778},
	title = {Deep Residual Learning for Image Recognition},
	year = {2016}}

@inproceedings{rom_dupont2019ANODE,
	author = {Dupont, Emilien and Doucet, Arnaud and Teh, Yee Whye},
	booktitle = {Advances in Neural Information Processing Systems},
	editor = {H. Wallach and H. Larochelle and A. Beygelzimer and F. d\textquotesingle Alch\'{e}-Buc and E. Fox and R. Garnett},
	publisher = {Curran Associates, Inc.},
	title = {Augmented Neural {ODEs}},
	url = {https://proceedings.neurips.cc/paper/2019/file/21be9a4bd4f81549a9d1d241981cec3c-Paper.pdf},
	volume = {32},
	year = {2019}}

@incollection{rom_hochreiter2001,
	author = {Hochreiter, S. and Bengio, Y. and Frasconi, P. and Schmidhuber, J.},
	booktitle = {A Field Guide to Dynamical Recurrent Neural Networks},
	description = {idsia},
	editor = {Kremer, S. C. and Kolen, J. F.},
	publisher = {IEEE Press},
	title = {Gradient flow in recurrent nets: the difficulty of learning long-term dependencies},
	year = 2001}

@book{rom_goodfellow16DL,
	author = {Ian J. Goodfellow and Yoshua Bengio and Aaron Courville},
	publisher = {MIT Press},
	title = {Deep Learning},
	year = {2016}}

@book{rom_bennerMOR1,
	doi = {10.1515/9783110498967},
	editor = {Peter Benner and Stefano Grivet-Talocia and Alfio Quarteroni and Gianluigi Rozza and Wil Schilders and Lu{\'\i}s Miguel Silveira},
	isbn = {9783110498967},
	lastchecked = {2023-01-07},
	publisher = {De Gruyter},
	title = {Model Order Reduction -- {System-} and Data-Driven Methods and Algorithms},
	url = {https://doi.org/10.1515/9783110498967},
	year = {2021}}

@article{diss_ifset22,
	author = {Maximilian Kannapinn and Minh Khang Pham and Michael Sch{\"a}fer},
	doi = {10.1016/j.ifset.2022.103143},
	issn = {1466-8564},
	journal = {Innovative Food Science \& Emerging Technologies},
	pages = {103143},
	title = {Physics-based digital twins for autonomous thermal food processing: {Efficient}, non-intrusive reduced-order modeling},
	url = {https://www.sciencedirect.com/science/article/pii/S1466856422002284},
	volume = {81},
	year = {2022}}

@article{sig_Talis2021,
	author = {Talis, Torben and Weigert, Joris and Esche, Erik and Repke, Jens-Uwe},
	doi = {10.1002/cite.202100109},
	journal = {Chemie Ingenieur Technik},
	number = {12},
	pages = {2097-2104},
	title = {Adaptive Sampling of Dynamic Systems for Generation of Fast and Accurate Surrogate Models},
	volume = {93},
	year = {2021}}

@article{sig_Heinz2017,
	author = {Tim Oliver Heinz and Oliver Nelles},
	doi = {10.1016/j.procs.2017.08.112},
	issn = {1877-0509},
	journal = {Procedia Computer Science},
	pages = {1054-1061},
	title = {Iterative Excitation Signal Design for Nonlinear Dynamic Black-Box Models},
	volume = {112},
	year = {2017}}

@techreport{mot_foodcode2017,
	author = {{U.S. Food and Drug Administration}},
	title = {Food code},
	url = {http://www.fda.gov/FoodCode},
	year = {2017}}

@article{mpc_Alonso2021,
	author = {A.A. Alonso and J.L. Pitarch and L.T. Antelo and C. Vilas},
	doi = {10.1016/j.fbp.2021.02.013},
	issn = {0960-3085},
	journal = {Food and Bioproducts Processing},
	pages = {162-173},
	title = {Event-based dynamic optimization for food thermal processing: {High-quality} food production under raw material variability},
	volume = {127},
	year = {2021}}

@article{dt_prawiranto2021,
	author = {Prawiranto, Kevin and Carmeliet, Jan and Defraeye, Thijs},
	doi = {10.3389/fsufs.2020.606845},
	issn = {2571-581X},
	journal = {Frontiers in Sustainable Food Systems},
	title = {Physics-Based Digital Twin Identifies Trade-Offs Between Drying Time, Fruit Quality, and Energy Use for Solar Drying},
	volume = {4},
	year = {2021}}

@article{dt_shoji2022,
	author = {Kanaha Shoji and Seraina Schudel and Daniel Onwude and Chandrima Shrivastava and Thijs Defraeye},
	doi = {10.1016/j.resconrec.2021.105914},
	issn = {0921-3449},
	journal = {Resources, Conservation and Recycling},
	pages = {105914},
	title = {Mapping the postharvest life of imported fruits from packhouse to retail stores using physics-based digital twins},
	volume = {176},
	year = {2022}}

@article{dt_tagliavini2019,
	author = {G. Tagliavini and T. Defraeye and J. Carmeliet},
	doi = {10.1016/j.fbp.2019.07.013},
	issn = {0960-3085},
	journal = {Food and Bioproducts Processing},
	pages = {310-320},
	title = {Multiphysics modeling of convective cooling of non-spherical, multi-material fruit to unveil its quality evolution throughout the cold chain},
	volume = {117},
	year = {2019}}

@article{dt_defraeye2019,
	author = {Thijs Defraeye and Giorgia Tagliavini and Wentao Wu and Kevin Prawiranto and Seraina Schudel and Mekdim {Assefa Kerisima} and Pieter Verboven and Andreas B{\"u}hlmann},
	doi = {10.1016/j.resconrec.2019.06.002},
	issn = {0921-3449},
	journal = {Resources, Conservation and Recycling},
	pages = {778-794},
	title = {Digital twins probe into food cooling and biochemical quality changes for reducing losses in refrigerated supply chains},
	volume = {149},
	year = {2019}}

@techreport{dt_ISO23247,
	author = {{International Organization for Standardization}},
	key = {ISO 23247-1},
	title = {{Automation systems and integration -- Digital twin framework for manufacturing -- Part 1: Overview and general principles}},
	type = {ISO 23247-1},
	volume = {2021},
	year = {2021}}

@inbook{dt_stark2019,
	author = {Stark, Rainer and Damerau, Thomas},
	booktitle = {CIRP Encyclopedia of Production Engineering},
	doi = {10.1007/978-3-642-35950-7_16870-1},
	editor = {Chatti, Sami and Tolio, Tullio},
	isbn = {978-3-642-35950-7},
	publisher = {Springer},
	title = {{CIRP} Encyclopedia of Production Engineering -- {Digital} Twin},
	year = {2019}}

@article{dt_henrichs2022,
	article-number = {115},
	author = {Henrichs, Elia and Noack, Tanja and Pinzon Piedrahita, Ana Mar{\'\i}a and Salem, Mar{\'\i}a Alejandra and Stolz, Johnathan and Krupitzer, Christian},
	doi = {10.3390/s22010115},
	issn = {1424-8220},
	journal = {Sensors},
	number = {1},
	pubmedid = {35009655},
	title = {Can a Byte Improve Our Bite? {An} Analysis of Digital Twins in the Food Industry},
	volume = {22},
	year = {2022}}

@article{rom_khan2022,
	author = {Khan, Md. Imran H. and Sablani, Shyam S. and Nayak, Richi and Gu, Yuantong},
	doi = {10.1111/1541-4337.12912},
	journal = {Comprehensive Reviews in Food Science and Food Safety},
	number = {2},
	pages = {1409-1438},
	title = {Machine learning-based modeling in food processing applications: {State} of the art},
	volume = {21},
	year = {2022}}

@article{mpc_Alonso2013,
	author = {Antonio A. Alonso and Ana Arias-M{\'e}ndez and Eva Balsa-Canto and M{\'\i}riam R. Garc{\'\i}a and Juan I. Molina and Carlos Vilas and Marcos Villaf{\'\i}n},
	doi = {10.1016/j.foodcont.2013.01.002},
	issn = {0956-7135},
	journal = {Food Control},
	number = {2},
	pages = {392-403},
	title = {Real time optimization for quality control of batch thermal sterilization of prepackaged foods},
	volume = {32},
	year = {2013}}

@article{mpc_Huang1998,
	address = {St. Joseph, MI},
	author = {Huang, Y. and D. Whittaker, A. and E. Lacey, R.},
	doi = {10.13031/2013.17282},
	isbn = {0001-2351},
	journal = {Transactions of the ASAE},
	journal1 = {Trans. ASAE},
	journal2 = {Transactions of the ASAE},
	journal3 = {Transactions of the ASAE},
	number = {5},
	pages = {1511--1517},
	publisher = {ASAE},
	title = {Neural Network Prediction Modeling for a Continuous, Snack Food Frying Process},
	volume = {41},
	year = {1998}}

@article{mpc_Li2016,
	author = {Jianshuo Li and Qingyu Xiong and Kai Wang and Xin Shi and Shan Liang},
	doi = {10.1080/07373937.2015.1122612},
	journal = {Drying Technology},
	number = {12},
	pages = {1434-1444},
	publisher = {Taylor & Francis},
	title = {A recurrent self-evolving fuzzy neural network predictive control for microwave drying process},
	volume = {34},
	year = {2016}}

@article{dt_Lu2020,
	author = {Yuqian Lu and Chao Liu and Kevin I-Kai Wang and Huiyue Huang and Xun Xu},
	doi = {10.1016/j.rcim.2019.101837},
	issn = {0736-5845},
	journal = {Robotics and Computer-Integrated Manufacturing},
	pages = {101837},
	title = {Digital Twin-driven smart manufacturing: {Connotation}, reference model, applications and research issues},
	volume = {61},
	year = {2020}}

@inproceedings{sig_Gringard2016,
	author = {Gringard, Matthias and Kroll, Andreas},
	booktitle = {2016 IEEE Symposium Series on Computational Intelligence (SSCI)},
	doi = {10.1109/SSCI.2016.7849984},
	pages = {1-8},
	title = {On the parametrization of {APRBS} and multisine test signals for the identification of nonlinear dynamic {TS}-models},
	year = {2016}}

@article{sig_Heinz2018,
	author = {Tim Oliver Heinz and Oliver Nelles},
	doi = {10.1515/auto-2018-0027},
	journal = {at - Automatisierungstechnik},
	number = {9},
	pages = {714--724},
	publisher = {Public Library of Science},
	title = {Excitation signal design for nonlinear dynamic systems with multiple inputs -- {A} data distribution approach},
	volume = {66},
	year = {2018}}

@article{dt_Rosen2015,
	author = {Roland Rosen and Georg {von Wichert} and George Lo and Kurt D. Bettenhausen},
	issn = {2405-8963},
	journal = {IFAC-PapersOnLine},
	note = {15th IFAC Symposium on Information Control Problems in Manufacturing},
	number = {3},
	pages = {567-572},
	title = {About The Importance of Autonomy and Digital Twins for the Future of Manufacturing},
	volume = {48},
	year = {2015}}

@article{dt_Niederer2021,
	author = {Niederer, Steven A. and Sacks, Michael S. and Girolami, Mark and Willcox, Karen},
	date = {2021/05/01},
	doi = {10.1038/s43588-021-00072-5},
	id = {Niederer2021},
	isbn = {2662-8457},
	journal = {Nature Computational Science},
	number = {5},
	pages = {313--320},
	title = {Scaling digital twins from the artisanal to the industrial},
	volume = {1},
	year = {2021}}

@article{dt_Rasheed2020,
	author = {Rasheed, Adil and San, Omer and Kvamsdal, Trond},
	doi = {10.1109/access.2020.2970143},
	journal = {IEEE Access},
	pages = {21980-22012},
	title = {Digital Twin: {Values}, Challenges and Enablers From a Modeling Perspective},
	volume = {8},
	year = {2020}}

@techreport{dt_AIAA2020,
	author = {{AIAA Digital Engineering Integration Committee}},
	institution = {{American Institute of Aeronautics and Astronautics}},
	title = {Digital Twin: {D}efinition \& Value. {An} {AIAA} and {AIA} Position Paper},
	year = {2020}}

@inbook{dt_Grieves2017,
	author = {Grieves, Michael and Vickers, John},
	booktitle = {Transdisciplinary Perspectives on Complex Systems: New Findings and Approaches},
	doi = {10.1007/978-3-319-38756-7_4},
	editor = {Kahlen, Franz-Josef and Flumerfelt, Shannon and Alves, Anabela},
	isbn = {978-3-319-38756-7},
	pages = {85--113},
	publisher = {Springer International Publishing},
	title = {Digital Twin: {Mitigating} Unpredictable, Undesirable Emergent Behavior in Complex Systems},
	year = {2017}}

@article{rom_calka2021,
	author = {Maxime Calka and Pascal Perrier and Jacques Ohayon and Christelle Grivot-Boichon and Michel Rochette and Yohan Payan},
	doi = {10.1016/j.cmpb.2020.105786},
	issn = {0169-2607},
	journal = {Computer Methods and Programs in Biomedicine},
	pages = {105786},
	title = {Machine-Learning based model order reduction of a biomechanical model of the human tongue},
	volume = {198},
	year = {2021}}

@unpublished{rom_DynROMPatent,
	author = {{U.S. Provisional Patent Application No. 62/773 555}},
	month = {November},
	title = {Systems and Methods for Building Dynamic Reduced Order Physical Models},
	year = {2018}}

@article{dt_tao2019,
	author = {Fei Tao and Qinglin Qi},
	doi = {10.1038/d41586-019-02849-1},
	journal = {Nature},
	pages = {490-491},
	title = {Make more digital twins},
	volume = {573},
	year = {2019}}

@book{rom_Nelles2020,
	author = {Oliver Nelles},
	edition = {2nd},
	publisher = {Springer International Publishing},
	title = {Nonlinear System Identification -- {From} Classical Approaches to Neural Networks, Fuzzy Models, and Gaussian Processes},
	year = {2020}}

@article{dt_Verboven2020,
	author = {Pieter Verboven and Thijs Defraeye and Ashim K Datta and Bart Nicolai},
	doi = {10.1016/j.cofs.2020.03.002},
	issn = {2214-7993},
	journal = {Current Opinion in Food Science},
	pages = {79-87},
	title = {Digital twins of food process operations: {The} next step for food process models?},
	volume = {35},
	year = {2020}}

@electronic{rom_twinbuilder_2020,
	author = {{ANSYS Inc.} },
	title = {{ ANSYS Twin Builder} -- {Release} 2020 {R2}},
	url = {https://www.ansys.com/},
	urldate = {2022-12-04},
	year = {2020}}

@electronic{dt_fmu,
	author = {{Modelica Association}},
	title = {Functional Mock-up Interface 2.0.},
	url = {https://fmi-standard.org},
	urldate = {2023-02-24},
	year = {2017}}

@article{cm_feyissa_3D2013,
	author = {Aberham Hailu Feyissa and Krist V. Gernaey and Jens Adler-Nissen},
	doi = {10.1016/j.meatsci.2012.12.003},
	issn = {0309-1740},
	journal = {Meat Science},
	number = {4},
	pages = {810-820},
	title = {{3D} modelling of coupled mass and heat transfer of a convection-oven roasting process},
	volume = {93},
	year = {2013}}

@article{cm_rabeler_mod2018,
	author = {Felix Rabeler and Aberham Hailu Feyissa},
	doi = {10.1016/j.jfoodeng.2018.05.021},
	issn = {0260-8774},
	journal = {Journal of Food Engineering},
	pages = {60-68},
	title = {Modelling the transport phenomena and texture changes of chicken breast meat during the roasting in a convective oven},
	volume = {237},
	year = {2018}}

@inbook{dt_Glaessgen2012,
	author = {Glaessgen, Edward and Stargel, David},
	booktitle = {53rd AIAA/ASME/ASCE/AHS/ASC Structures, Structural Dynamics and Materials Conference},
	doi = {10.2514/6.2012-1818},
	m1 = {0},
	m3 = {10.2514/6.2012-1818},
	publisher = {American Institute of Aeronautics and Astronautics},
	title = {The Digital Twin Paradigm for Future {NASA} and {U.S.} Air Force Vehicles},
	title1 = {Structures, Structural Dynamics, and Materials and Co-located Conferences},
	ty = {CHAP},
	year = {2012},
	year1 = {2012/04/23}}

@article{valizadeh2023,
  title = {Influence of Process Parameters on Geometric and Elasto-Visco-Plastic Material Properties in Vat Photopolymerization},
  author = {Valizadeh, Iman and Tayyarian, Tannaz and Weeger, Oliver},
  date = {2023-06-05},
  journaltitle = {Additive Manufacturing},
  shortjournal = {Additive Manufacturing},
  volume = {72},
  pages = {103641},
  issn = {2214-8604},
  doi = {10.1016/j.addma.2023.103641},
}

%
%
%
%
\end{document}

\endinput